\documentclass[12pt]{article}
\usepackage{graphicx}
\usepackage{cite}
\usepackage{amsmath}

\begin{document}

\title{Unanimity, Coexistence and Rigidity: Three Sides of Polarization }

\author{ Serge Galam\thanks{serge.galam@sciencespo.fr} \\
CEVIPOF - Centre for Political Research, Sciences Po and CNRS,\\
1, Place Saint Thomas d'Aquin, Paris 75007, France}

\date{March 6, 2023}

% S. Galam, Unanimity, Coexistence, and Rigidity: Three Sides of Polarization. Entropy 2023, 25, 622

\maketitle

\begin{abstract}

Political polarization is perceived as a threat to democracies. Using the Galam model of opinion dynamics deployed in a five-dimensional parameter space, I show that polarization is the byproduct of an essential hallmark of a vibrant democratic society, namely the open and informal discussions among agents. Indeed, within a homogeneous social community with floaters, the dynamics leads gradually toward unanimity (zero entropy). Polarization can eventually appear as the juxtaposition of non-mixing social groups sharing different prejudices about the issue at stake. On the other hand, the inclusion of contrarian agents produces a polarization within a community that mixes when their proportion $x$ is beyond a critical value $x_c=\frac{1}{6}\approx 0.167$ for discussing groups of size 3 and 4. Similarly, the presence of stubborn agents produces also a polarization of a community that mixes when the proportion of stubborns is greater than some critical value. For equal proportions of stubborns $a$ along each opinion, $a_c=\frac{2}{9}\approx 0.22$ for group size 4 against $a_c=\frac{1}{4}= 0.25$ for group size 3. However, the evaluation of the proportion of individual opinion shifts at the attractor $\frac{1}{2}$ indicates that the polarization produced by contrarians is fluid  with a good deal of agents who keep shifting between the two opposed blocks (high entropy). That favors a coexistence of opposite opinions in a divided community. In contrast, the polarization created by stubborn agents is found to be frozen with very few individuals shifting opinion between the two opinions (low entropy). That yields a basis for the emergence of hate between the frozen opposed blocks.

\end{abstract}


\section{Introduction}

In the last years the phenomenon of political polarization has become an issue of major concern among scholars, pundits, journalists and politicians  \cite{p1, p2, p3, p4, p5, pp6, pp7}. Indeed, the current polarization of modern societies is often perceived as a direct and immediate threat to the stability of democratic societies.  Both 2021 American  \cite{pa} and 2023 Brazilian  \cite{pb} elections as well as the Israeli 2023 crisis related to the will of reforming the judicial system  \cite{pc}, have enlightened the reality of this fear by exhibiting countries divided into two parts almost equal, opposite, irreconcilable and hating each other.  When the hateful trait is present, the polarization is often referred to as affective polarization \cite{p3, paff2}. Polarization also emerges on non-political but societal issues or challenges like global warming, Brexit, nuclear energy, secularism.

While the issue of understanding the phenomenon of polarization has attracted a great deal of works, there is still no consensus among the researchers working on the topic as to what causes polarization to occur in a given population polarization  \cite{p6, p7, p8}. A good part of the works addresses the issue from the perspective of the dynamics of opinion within sociophysics  \cite{brazil, frank, book, bikas},  Most of related papers consider binary variables  \cite{nun1, nun2, sen, tot, mala, and, red, gm3, kas1, bol, che, mau, nun3, bag, car,  kas2, zan, flo, mar, igl, fas, gim, iac, bru} with a few ones having three discrete opinions  \cite{celia, mobilla}. Among the numerous models is the Galam model, which has been deployed in a multi-dimensional space of parameters to study the competition between discrete opinions  \cite{gtak, uni2, vot, chop, min, het, cont, inf, pair, gmar}. 

In this paper, I review and extend the Galam model of opinion dynamics in connection to the emergence of polarization. Introducing a novel quantify which calculates the proportions of agents shifting opinion at a given moment, I found that polarization is a  threefold feature, which I denote unanimity, coexistence and rigidity.

The origin of each side of the opinion update scheme is found to stem from heterogeneity among the psychological characters of the agents composing the social group when engaged in a debate about a collective issue. Up to date, three particular psychological traits have been included with the floaters, the contrarians and the stubborns.

\begin{description}

\item[The floaters:]  floaters are agents who have an opinion, argue and vote for it.  But eventually they may shift to the opposite one when being minority in a local discussing group  \cite{vot, chop, min, het}.

\item[The contrarians:] contrarians are agents who have an opinion, argue vote for it but eventually shift to the opposite one when being majority in their local discussing group. The shift is independent of the opinions themselves \cite{cont}.

\item[The stubborns:]  stubborns are agents who do have an opinion, argue and vote for it. But contrary to other agents, they stick to their initial opinions whatever is the composition of their local discussing group \cite{inf, pair, gmar}.

\end{description}

The various associated effects on the opinion dynamics can be investigated thanks to a universal formula for the updates of opinions. The formula has been derived within a parameter space of five dimensions.  which are the size of the discussing groups, the proportion of contrarians, the two proportions of stubborns for respective opinions, and the distribution of prejudices in favor of each opinions \cite{gtak}.

The corresponding multi-dimensional phase diagram is rather rich with a combination of both tipping point dynamics and single attractor dynamics as a function of the values of the five parameters. Polarization, unanimity and  coexistence are then recovered as respective deviations from one another as a function of the respective proportions of each psychological subgroup. Those deviations are pined by the psychological traits which deviate the associated dynamics towards unanimity. 

New results are obtained with respect to the nature of a polarizated stable state. In particular floaters are found to produce segregated polarization (zero entropy), contrarians a fluid polarization (high entropy) and stubborns a frozen polarization (low entropy).

The rest of the paper is organized as follows: Section 2 sets the ground of opinion dynamics with the various ingredients used in the paper. The spontaneous drive towards democratic unanimity is outlined in Section 3. The prejudices breaking of a perfect democratic dynamics is review in Section 4 while Section 5 investigate the effect of having contrarian agents on the related dynamics of opinion. The stubborn agents are introduced in Section 6 followed by some conclusion.

\section{Opinion dynamics, definitions and reality}

The phenomenon of polarization is mainly used to describe large communities of agents which, over some period of time, get divided into two opposite groups, each considering the other as extreme. These extreme societal visions are generally deployed along a global societal project.

To address this phenomenon and account for above definition I consider a community of people who have to decide between two opposite choices denoted A and B. Before the launching of the collective campaign, each agent has reached a choice either A or B according to its own values, experiences, visions. These individual alignments yield initial proportions $p_0$ and $(1-p_0)$ of agents holding respectively opinions A and B. What made each agent initial choice is out the scope of the present work. I only assume that agents are aware of what motivates their respective choices and will argue to promote it among those who have made the opposite choice when debating the topic in a social informal meeting.

Moreover, the model assumes that people discuss  informally the issue at stake in small groups during social events, like dinners, lunches, coffees, drinks, commuting and more. Even, during large gatherings of people like for instance at a weeding, the assembly divides into small groups.These on going and repeated encounters in small groups shape individual opinions, which end up with time into aggregated collective opinions.

The debate will eventually modify the proportions $p_0$ and $(1-p_0)$ to new values $p_T$ and $(1-p_T)$ where T is a function of the time at which either a vote is taking place or people stop debating the topic to eventually start on a new one. On this basis I define three states of polarization.

\begin{enumerate}
\item Unanimity: when $p_T$ is either equal to 1 or very large around 0.80 as well as equal to 0 or very low around 0.20, I define the associated state as unanimity. Most agents share opinion A in the first case and opinion B in the second one. The values 0.80 and 0.20 are chosen arbitrarily to set a boundary beyond which one opinion overwhelms the other. In reality these values fluctuate but preserve the feeling of a landslide victory. Having an overwhelming majority of agents who share the same opinion against a small minority holding the other opinion makes the related entropy to be small and even zero in cases $p_T=1$ and $p_T=0$.

\item Coexistence: when $p_T$ is of the order of $0.50 \pm 0.03$ i. e., 0.53 and 0.47. I define the associated state as coexistence if and only if a substantial part of the population keep shifting opinion without modifying the overall proportions $p_T$ and $(1-p_T)$.  It means that the global opinion has reached an attractor located around 0.50 but individual choices are not frozen with noticeable parts of the agents who keep shifting opinions. The value $\pm 0.03$ is chosen arbitrarily to set a fuzzy boundary around 0.50. In reality these values fluctuate a bit but preserve the feeling of a hung outcome in case of an election. The high level of ongoing shift of individual opinion put the associated entropy at a high value.

\item Rigidity: when $p_T$ is of the order of $0.50 \pm 0.03$ i. e., 0.53 and 0.47, I define the associated state as rigidity, if and only if the stable global opinion around 0.50 is frozen at individual choices. No noticeable part of the agents keep shifting opinions. The choices of $\pm 0.03$ is chosen arbitrarily to set a boundary around 0.50. In reality these values fluctuate a little while preserving the feeling that the winner has stolen its victory to the competitor in the event of an election. In this case, the low level of individual opinion shifts produces a low entropy.

\end{enumerate}

At this stage, to avoid any misunderstanding, I would like to emphasize that the results obtained from the model, should not be taken literally. They are intended to be indicators of the hidden trends, which drive the dynamics of the social and political reality,   the model aims to describe.

\section{The spontaneous drive towards democratic unanimity}

The Galam model of opinion dynamics operates in three successive steps, which are repeated a number T of times. First, agents are distributed randomly in small groups. Second, majority rules are applied simultaneously in each group to update locally the opinions of agents. Third, all agents are reshuffled.

When the population is homogeneous being composed of only floaters, all agents who are minority in a group shift opinion to adopt the one having gained the vote majority \cite{vot, chop, min}.

For the sake of readiness and analytical solving I restrain the review of update groups of size  $r=4$. A random distribution of A and B agents in a group of four leads to $2^4=16$ possible configurations. Five configurations have a majority of A, five a majority of B and 6 have 2 A and  2 B. Majority rule attributes the first five to A and the last five to B. The six configurations with a tie have not majority and a physicist would assume that in this case no update occur keeping 2 A and 2 B. 

Accordingly, starting with an initial proportions $p_0$ and $(1-p_0)$ of agents holding respectively opinions A and B, one update cycle leads to new proportions $p_1$ and $(1-p_1)$ where $p_1$ is given by, 
\begin{equation}
p_{1}= p_0^4+4p_0^3(1-p_0)+3 p_0^2(1-p_0)^2 .
\label{p1} 
\end{equation}
First two terms of Eq. (\ref{p1}) accounts for a majority of A among the four agents yielding four A and last term accounts for the unchanged tie configurations 2A and 2B.

After one cycle of updates, agents are reshuffled and distributed again randomly in groups of size four leading to a proportion $p_2$ obtained from Eq. (\ref{p1}) applied to $p_1$. Repeating the process yields a tipping point dynamics as seen by solving the fixed point equation $p_1=p_0$, which yields two attractors $p_A=1$ and $p_B=0$ separated by a tipping point in between located at $p_{c}=\frac{1}{2}$.

At this stage, it is worth to notice that expanding  Eq. (\ref{p1}) yields $p_{1}=-2p^3+3p^2$, which is equal to $p_{1}=p^3+3p^2(1-p)$, which turns out to be the update Equation for groups of size three. Keeping invariant the tie configuration makes update groups of four agents identical to update groups of three.

The tipping point $p_{c}=\frac{1}{2}$ makes iteration of Eq. (\ref{p1}) to produce a series $p_0<p_1< \dots <p_T \rightarrow 1$ when $p_0>0.50$. For  $p_0<0.50$ the series is $p_0>p_1> \dots >p_T \rightarrow 0$. Therefore, the initial majority of aggregated opinions convinces the minority through local and repeated discussions leading towards unanimity, with all agents sharing the same opinion, provided enough updates have been made. In this case, at the attractor the related entropy is thus zero.

Provided that people keep discussing for a sufficient time, I get a representation of an ideal perfect democracy. Through informal and open mind discussions, conflicts between opinions have disappeared with no more difference among the opinions of agents.  No additional arguing is required with everyone sharing the same opinion via a rational process, which made the opinion initially supported by the majority of the agents prevail. The unanimity, which has emerged is thus democratic.

It is of importance to underline that in connection to reality, one cycle of opinion update driven by Eq. (\ref{p1}) is the equivalent of an average of several local discussions in the real world. The number of these real discussions is a matter of intensity of the ongoing campaign. 

\section{Prejudices break unconsciously the perfect democratic dynamics}

However, even within an ideal society, perfection does not exist. Paradoxically, rationality may produce local collective doubts by gathering arguments. The result is two different choices that seem equally valid, given the arguments for and against each. Indeed, such a situation appears quite naturally within the model in the tie configurations 2A-2B. 

In case of a tie, the agents thus decide either A or B, by chance, as in a coin toss. No rational argument is evoked in the selected choice. They could have equally selected the other one. Mathematically this tie breaking yields an identical contribution $3p_0^2(1-p_0)^2$ to Eq. (\ref{p1}). However, now at a tie, instead of no update implemented with 2A-2B, all four agents choose either A (4A) or B (4B) with equal probabilities. 

At this point, I introduce a fundamental assumption to incorporate the human character of agents, as opposed to the processing of atoms. I assume that in the event of a tie, all four agents doubt and that this state of doubt unconsciously opens the door to an invisible bias that will guide their choice.  Therefore, when the group selects ``by chance" one choice over the other, the "chance" is being biased along the leading prejudice, which is activated by the issue at stake. 

I account for that effect by allocating the group choice to A with probability $k$ and to B with probability $(1-k)$ \cite{het}. The value of $k$ is a function of distribution of prejudices, which are in tune with opinion A among the agents. The associated update Eq. (\ref{p1}) becomes, 
\begin{equation}
p_{1,k}= p_0^4+4p_0^3(1-p_0)+6k p_0^2(1-p_0)^2 ,
\label{p1k} 
\end{equation}
which yields the same two attractors $p_{A,k}=1$ and $p_{B,k}=0$ as above but now the tipping point is located at,
\begin{equation}
p_{c,k}=\frac{(6k-5)+\sqrt{13-36k+36k^2}} {6(2k-1)} .
\label{pck} 
\end{equation}

Eq. (\ref{pck})  gives $p_{c,0}=\frac{5-\sqrt{13}} {6} \approx 0.23$,  $p_{c,\frac{1}{2}}=\frac{1}{2}$ and $p_{c,1} = \frac{1+\sqrt{13}} {6} \approx 0.77$ for respectively $k=0, \frac{1}{2}, 1$. Therefore $0\leq k \leq \frac{1}{2} \Rightarrow 0.23\  \leq p_{c,k}\leq \frac{1}{2}$  and $\frac{1}{2}\leq  k \leq 1 \Rightarrow \frac{1}{2} \leq p_{c,k}\leq  0.77$.

The case $k=1$ illustrates the phenomenon of minority spreading. Opinion A being favored by the group prejudice, it needs to gather an initial minority proportion of only 0.23 to convince the initial majority of agents who are sharing opinion B to adopt instead opinion A via local and open mind discussions. The previous democratic character the dynamics has been broken naturally and unconsciously without notice. No coercion has been used.

\subsection{Segregated polarization}

While the prejudice effect preserves the dynamic of unanimity, it produces a democratic breaking at the advantage of the opinion in tune with the leading prejudice of a social community. This opinion is either initially held by a minority or a majority of agents. 

However, within the same country different social communities are spread over with their respective members not mixing together in social meetings where informal discussions are being held. And it often happens that in these different communities different prejudices are activated by the same issue at stake.  As a result, opposite opinions may end up to spread along opposite unanimities in adjacent communities creating a de facto stable polarization at a higher aggregated level.

Nevertheless, this segregated frozen polarization is not perceived as a threat or a problem since it subscribes to the specific features, which make what differentiates those communities. Moreover, it is usually not used as a background to reach power at global levels, which overrules all communities.

\subsection{Combining group of different sizes}

I have restricted the update Equations to groups of size three and four but in practice people discuss in group of different sizes $s$. On this basis, it is possible to consider a distribution of group sizes from $s=1$ to $s=L$ where size 1 accounts for agents who do not discuss during one update and $L$ is the larger size of group discussion. L is rarely larger than five or six since informal larger groups always split spontaneously in smaller groups.

For a distribution of size $s$ with respective proportions $g_s$, Eq. (\ref{p1}) becomes,
\begin{equation}
p_{L,1}=\sum_{s=1}^{L} \left\{ g_{s} \left [ \sum_{l=\bar s+1}^{s}   {s \choose l} p_{0}^l  (1-p_0)^{s-l} +\delta [\bar s-\frac{s}{2}]k {s \choose s/2}  p_{0}^{\frac{s}{2}}  (1-p_{0})^{\frac{s}{2}} \right ] \right\}  \ ,
\label{pl} 
\end{equation}
where $\bar s \equiv [\frac{s}{2}]$ with $[\dots]$ meaning the integer part, $\delta [\dots]$ is the Kronecker function and,
\begin{equation}
\sum_{s=1}^{s=L}g_s=1 .
\label{gs} 
\end{equation}

The two attractors $p_{A,k} = 1$ and $p_{B,k = 0}$ obtained from Eq. (\ref{p1k}) are also the attractors associated to Eq. (\ref{pl}). However, the value of the tipping point $p_{c,k}$ is now modified as expected. On the one hand, larger even sizes reduce the probability of occurrence of a tie shifting $p_{c,k}$ closer to $\frac{1}{2}$, but on the other hand, for size 2 the tipping point is located at 0 for  $k=1$ and at 1 for $k=0$. These two opposite effects shows that having a combination of sizes for small groups will not modified qualitatively the results obtained for sizes " and 4. For instance, choosing $g_1=0.20, g_2=0.30, g_3=0.20, g_4=0.20, g_5=0.10$ yields $p_{c,0}=0.85$ and $p_{c,1}=0.15$ instead of $p_{c,0}=0.77$ and $p_{c,1}=0.23$ for size 4.

\section{Contrarians fuel coexistence}

In the case of no prejudice effect ($k=\frac{1}{2}$), the inclusion of a proportion $x$ of contrarians makes the equation (\ref{p1k}) to be written as \cite{cont}, 
\begin{equation}
p_{1,  x}= (1-2x) \left[ p_0^4+4p_0^3(1-p_0)+3 p_0^2(1-p_0)^2 \right] +x,
\label{p1x} 
\end{equation}
which yields three fixed points $p_{c,x}=\frac{1}{2}$ and,
\begin{equation}
p_{{A, x};{B,  x}}=\frac{-1 + 2 x \pm \sqrt{1 -8 x+12 x^2}}{2 (-1 + 2 x)} .
\label{pABx} 
\end{equation}
with last two being valid only in the range $0\leq x < x_c=\frac{1}{6} \approx 0.167$ and $x\geq \frac{1}{2}$. However, for $x\geq \frac{1}{2}$ the two values are not valid since there, $p_{{A, x}}>1$    and $p_{{B, x}}<0$.

\begin{figure}
\includegraphics[width=1\textwidth]{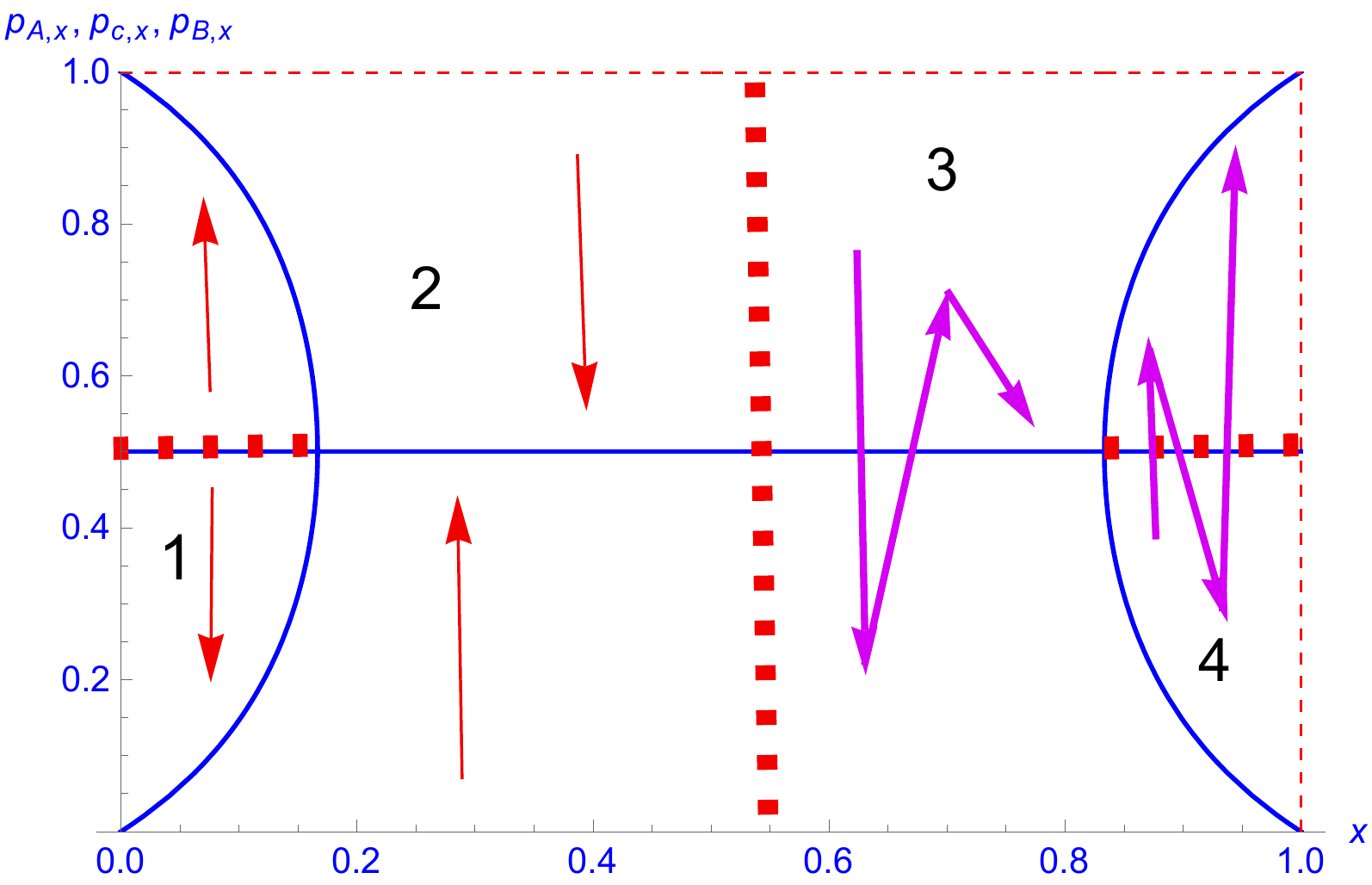}
\caption{Contrarians produce four distinct regions with different behaviors as a function of their proportion. A tipping point dynamics with $p_{c,x}=\frac{1}{2}$ prevails in the range $0\leq x < x_c$ (region 1). The two associated attractors feature a stable coexistence of a majority and a minority. In the range $x_c \leq x \leq \frac{1}{2}$ the dynamics turns into a one attractor dynamics located at $\frac{1}{2}$. Any initial support $p_0$ moves monotonously towards $\frac{1}{2}$ with repeating local updates (region 2). There, both opinions coexist in a perfect overall balance. When $\frac{1}{2}< x < \frac{5}{6}$ the dynamics is still monitored by one attractor at $\frac{1}{2}$ but the convergence towards it becomes oscillatory (region 3). The fourth region extends in the range $\frac{5}{6}< x \leq 1$. The dynamics gets back to a tipping point one but with an oscillatory dynamics between the two attractors (region 4).}
\label{r1}
\end{figure}

\begin{figure}
\includegraphics[width=.50\textwidth]{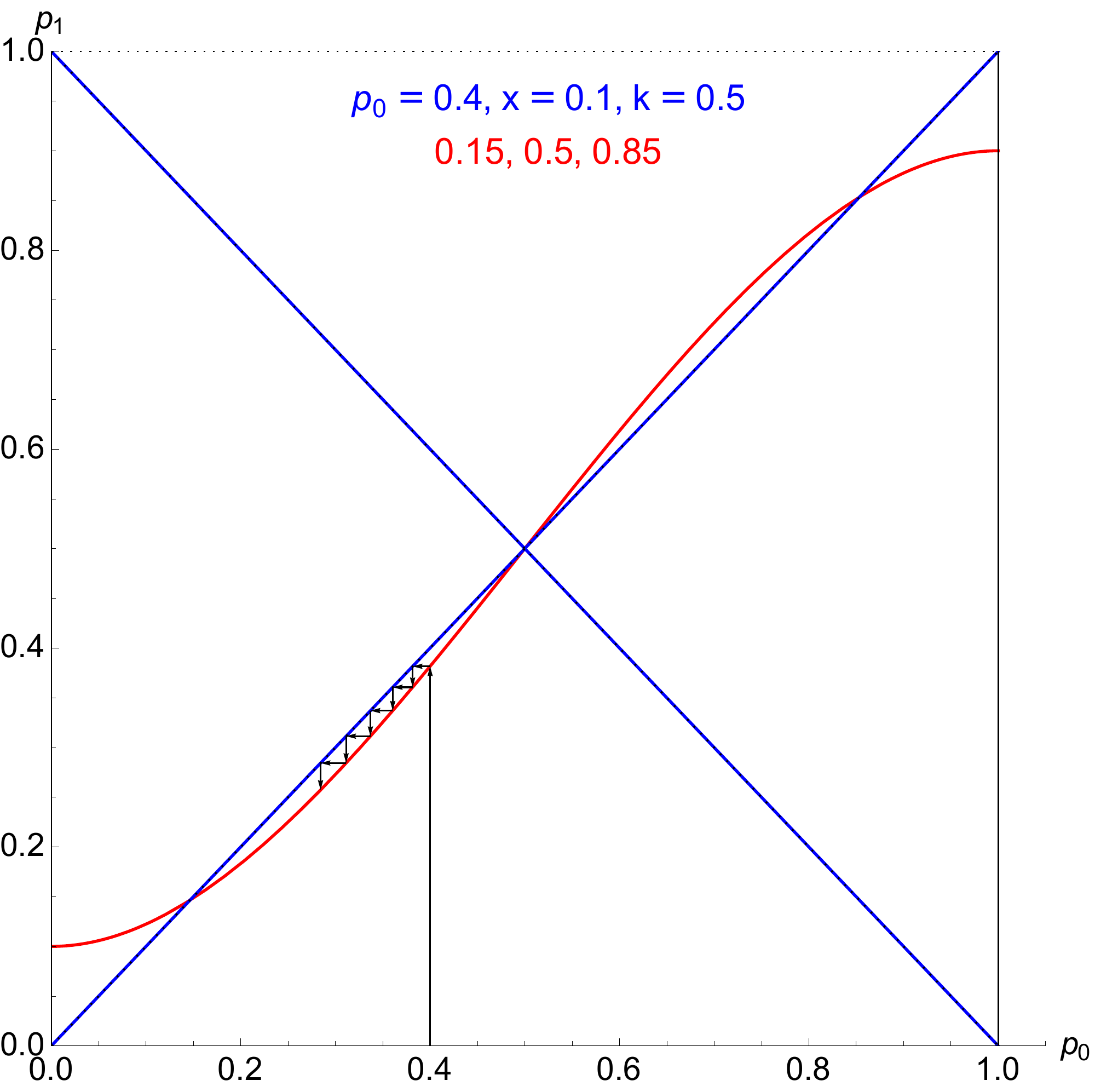}\quad
\includegraphics[width=.50\textwidth]{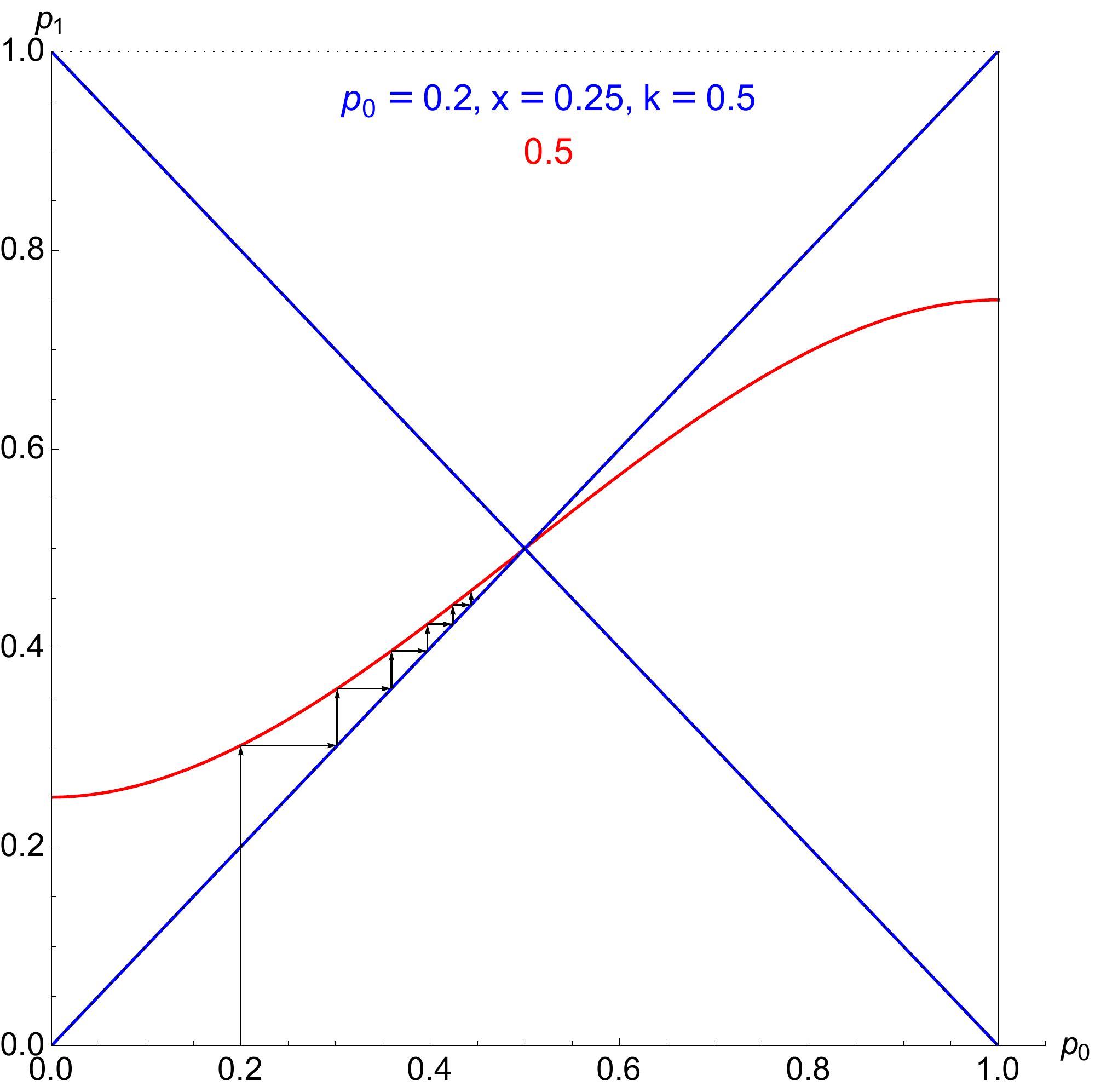}\\ \\ \\
\includegraphics[width=.50\textwidth]{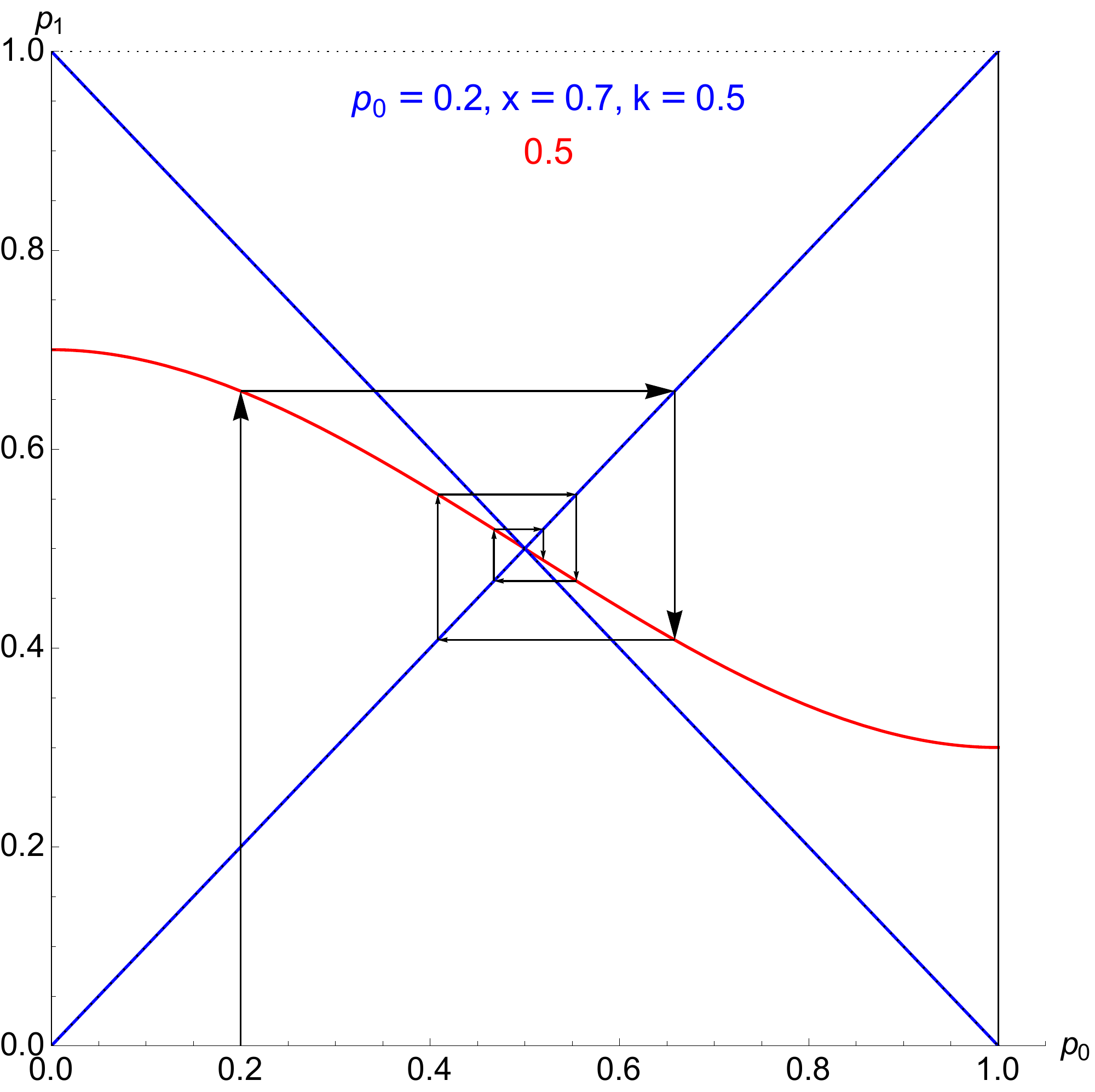}\quad
\includegraphics[width=.50\textwidth]{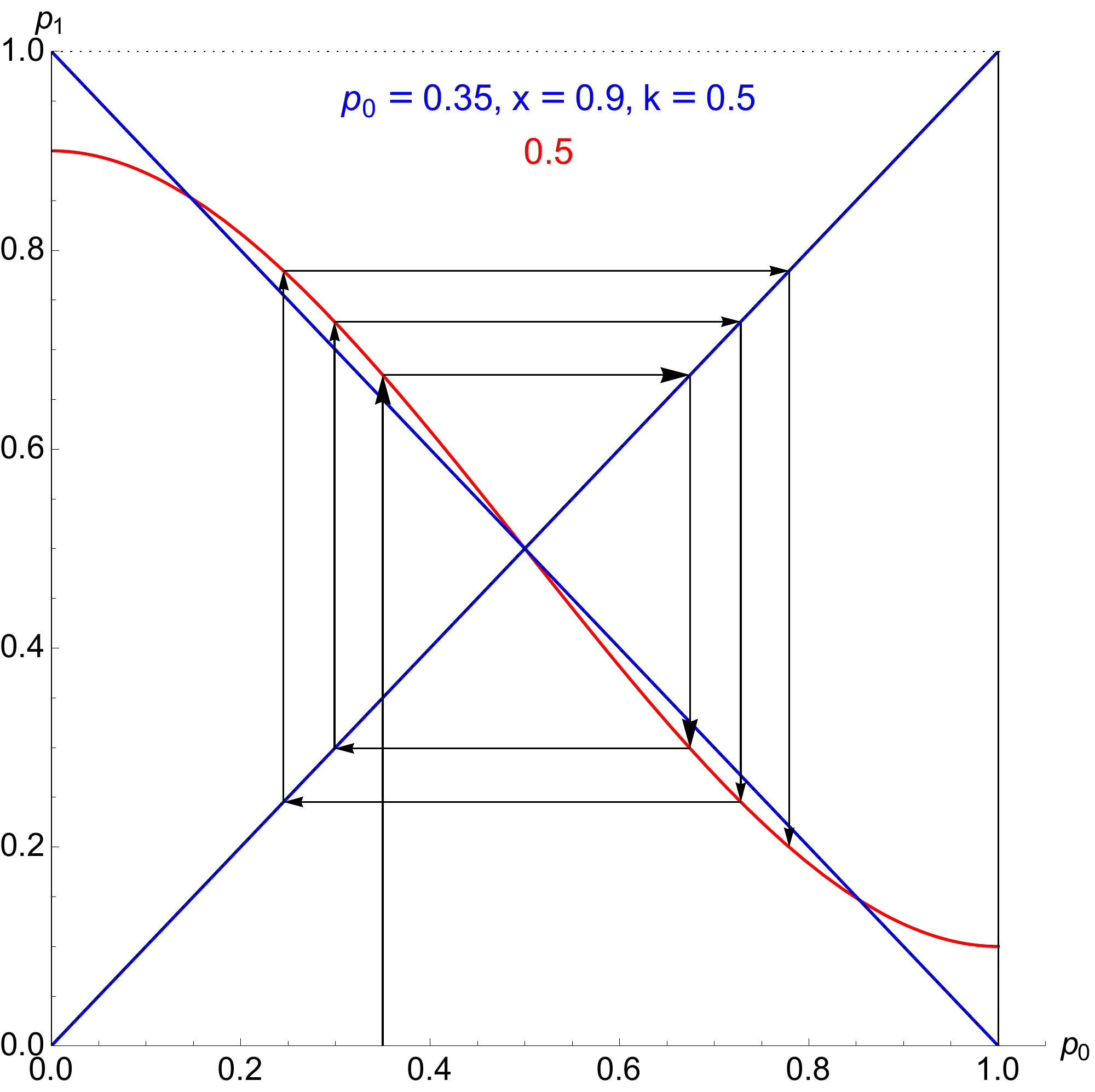}
\caption{Illustration of the dynamics of opinions in each one of the four regions produced by contrarians with respectively $x=0.10, p_0=0.40$ (upper left), $x=0.25, p_0=0.20$ (upper right), $x=0.70, p_0=0.20$ (lower left), $x=0.90, p_0=0.35$ (lower right). All cases have $k=\frac{1}{2}$. The respective values of the attractors and tipping points are indicated (in red).}
\label{r2}
\end{figure}

The associated dynamics is identified using the parameter $\lambda_x=\frac{p_{1, x}}{dp_0} \vert_{\frac{1}{2}}= \frac{3}{2}(1-2x)$ to determine the stability of $p_{c,x}$.
The fixed point  $p_{c,x}=\frac{1}{2}$ is thus a tipping point when $\lambda_x <-1$ and $\lambda_x >1$ making $p_{{A, x}; {B, x}}$ the two attractors of the dynamics. For $-1< \lambda_x < 1$, $p_{c,x}$ is an attractor, which implies that in this range $p_{{A, x}; {B,  x}}$ do not exist. Accordingly, to get the associated phase diagram I solve the inequality which makes  $p_{c,x}$ an attractor,
\begin{equation}
-1<  \frac{3}{2}(1-2x) < 1,
\label{sta1} 
\end{equation}
which is identical to, 
\begin{equation}
\frac{1}{6} <  x < \frac{5}{6}.
\label{sta2}
\end{equation}
Four distinct regions with different behaviors are obtained from Eq. (\ref{sta2}) as shown in Figs. (\ref{r1},\ref{r2}). 

\begin{description}
\item[Region 1] lies within the range $0\leq x < x_c$ featuring a tipping point dynamics with $p_{c,x}=\frac{1}{2}$ being the tipping point. The initial majority is increased by the repeated cycles of local discussions with a monotonic convergence towards the relevant attractor either $p_{{A, x}}$ when $p_0> \frac{1}{2}$ and $p_{{B, x}}$ when $p_0< \frac{1}{2}$. In the first case, A wins the public debate or the related vote but loses in the second case. In both cases, a core minority B (A) subsists against the majority A (B). The two attractors moves towards each other with increasing $x$ towards $x_c$. 

\item[Region 2] starts at $x_c$ where the two attractors $p_{{A, x}}$ and $p_{{B, x}}$ merge and disappear at $p_{c,x}$ turning the tipping point $p_{c,x}$ into an attractor. The dynamics shifts suddenly from a tipping point one to a single attractor dynamics. In the range $x_c \leq x \leq \frac{1}{2}$ any initial proportion $p_0$ is moved monotonously by the update dynamics towards $\frac{1}{2}$, i.e., an equal proportion of agents holding opinions A and B respectively. We have a perfect stable coexistence of both competing opinions in the range $x_c\leq x \leq \frac{1}{2}$.

\item[Region 3] marks the transition to a situation where contrarians are more numerous than floaters with $x>\frac{1}{2}$. Due to this fact, while $p_{c,x}=\frac{1}{2}$ remains an attractor, reaching it follows an oscillatory convergence. The oscillatory convergence holds in the range $\frac{1}{2}< x < \frac{5}{6}$. Once the attractor has been reached, the two competing opinions coexist at equal proportions as Region 2.  

\item[Region 4] is the counter part of region 1 where $p_{c,x}=\frac{1}{2}$ is again a tipping point. But now, the very high proportion of contrarians turn the dynamics into an oscillating divergence from the tipping point instead of a monotonic divergence. In addition, once an attractor has been reached, the dynamics becomes oscillating between  $p_{{A,  x}}$ and $p_{{B, x}}$. Region 4 extends in the range  $\frac{5}{6}< x \leq 1$.

\end{description}

\subsection{The polarization at $p_{c,x}=\frac{1}{2}$ is fluid}

In Regions 2 and 3 the stable state is a perfect equality between the respective numbers of agents holding opinions A and B. The community is thus divided into opposite parts, which in turn could lead to feature it as a polarized community. However, that could be misleading since the two opposite parts are not two frozen opposite parts. 

Indeed, contrarians make the division fluid with a good number of agents constantly moving from one side to the other but in equal proportions. I thus denote that fluid polarized state as coexistence.

To quantify the degree of fluidity of coexistence I introduce four new quantities,

\begin{eqnarray}
 S_{A, p, x_-}&=&(1 - x) \left[p^3 (1 - p) + \frac{3}{2} p^2 (1 - p)^2\right] , \nonumber\\
 S_{A, p,x_+}&=&x  \left[(1 - p)^4 + 3 p (1 - p)^3 + \frac{3}{2} p^2 (1 - p)^2\right] , \nonumber\\
S_{B, p,x_-}&=&(1 - x) \left[p (1 - p)^3 + \frac{3}{2}p^2 (1 - p)^2\right] , \nonumber\\
S_{B, p,x_+}&=&x  \left[p^4 + 3 p^3 (1 - p) + \frac{3}{2} p^2 (1 - p)^2\right] , \label{sab4}
\end{eqnarray}
which are, given proportions $p$ and $x$, the proportions of agents shifting opinions from B to A and from A to B. These shifts are  triggered respectively by local majorities  ($S_{A, p,x_-},  S_{B, p,x_-}$) and contrarians ($S_{A, p,x_+}, S_{B, p,x_+}$) during one update. 

The total proportion of agents shifting opinion from B to A during one update is thus $S_{A,p,x}=S_{A, p,x_-} +  S_{A, p,x_+}$ and $S_{B,p,x}=S_{B, p,x_-} +  S_{B, p,x_+}$ from A to B. The resulting total proportion of shifts is $S_{T, p, x}=S_{A, p, x} +  S_{A, p, x} $. With only floaters (no contrarians) the associated values are $S_{A, p,0}, S_{A, p,0}, S_{T,p,0}$ at $k=\frac{1}{2}$.

\begin{figure}
\includegraphics[width=.50\textwidth]{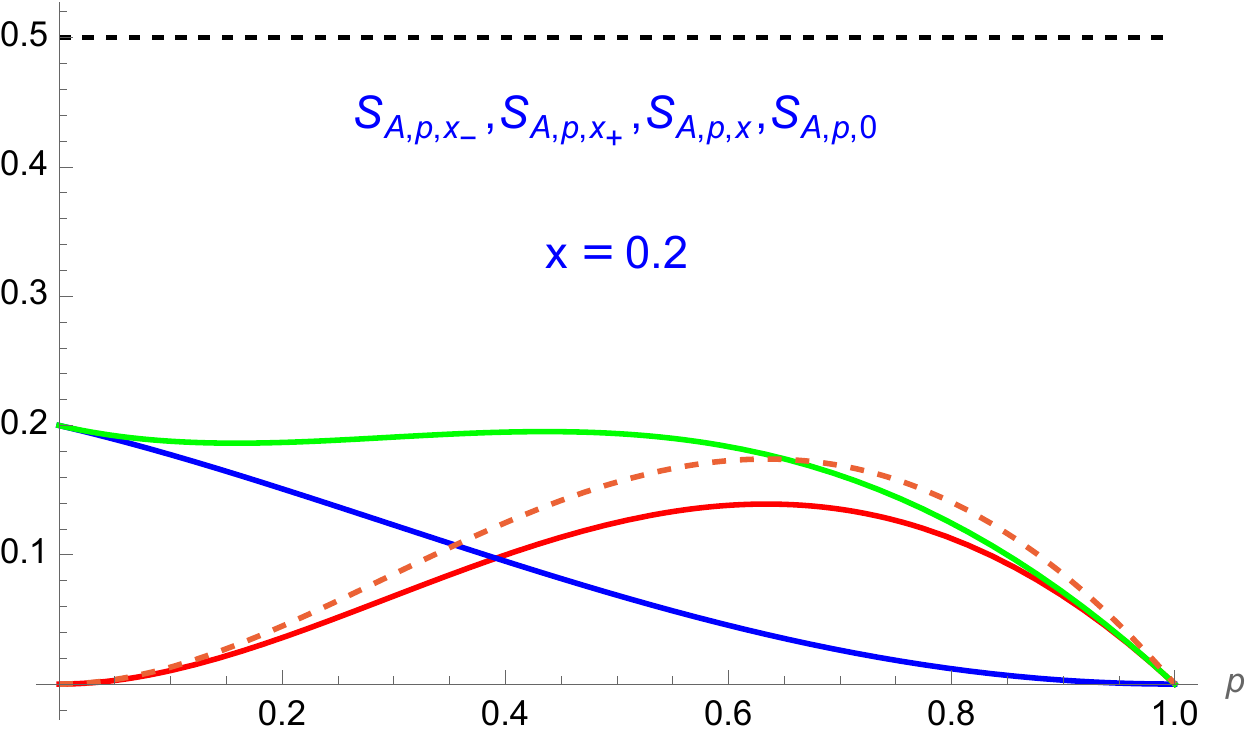}\quad
\includegraphics[width=.50\textwidth]{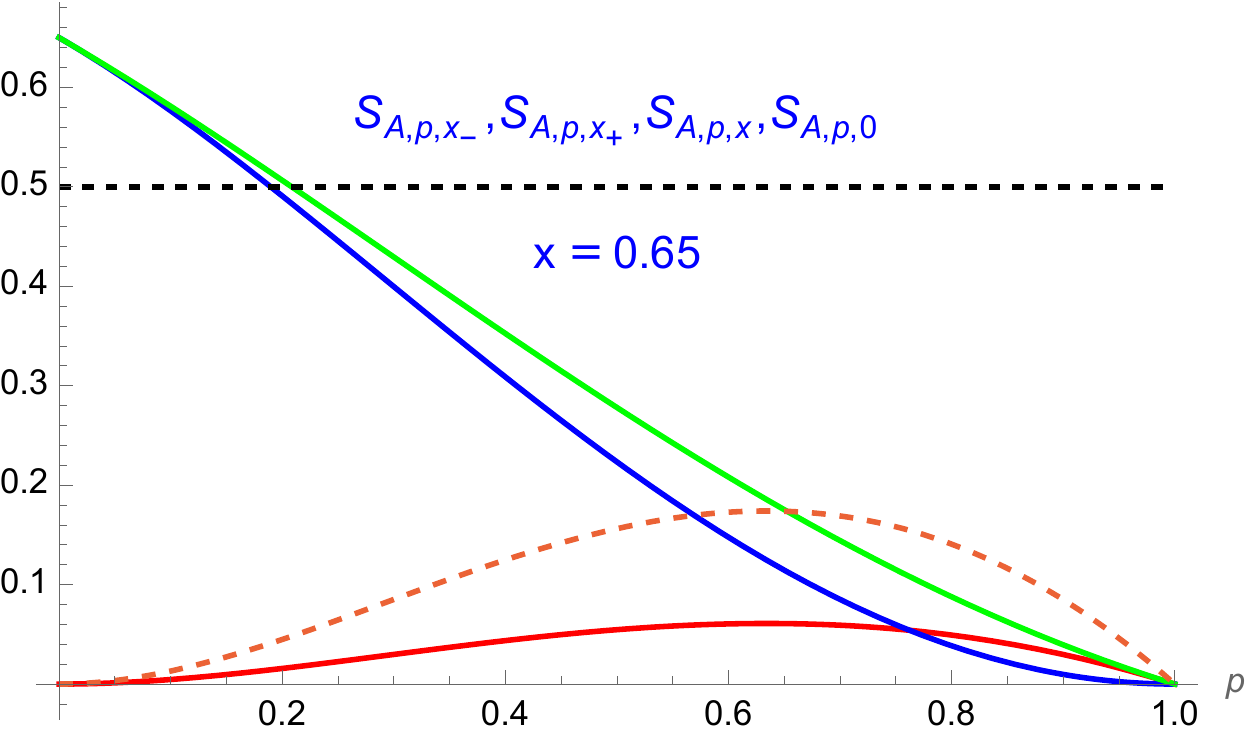}\\ \\ \\
\includegraphics[width=.50\textwidth]{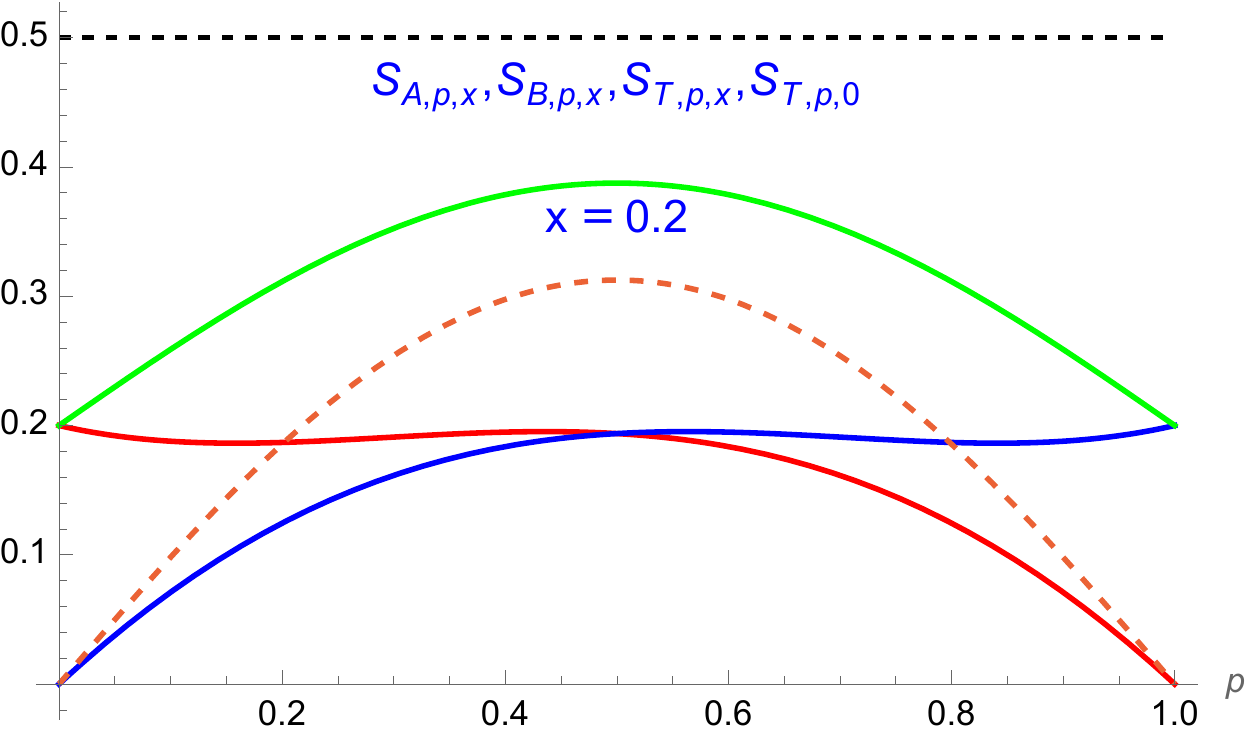}\quad
\includegraphics[width=.50\textwidth]{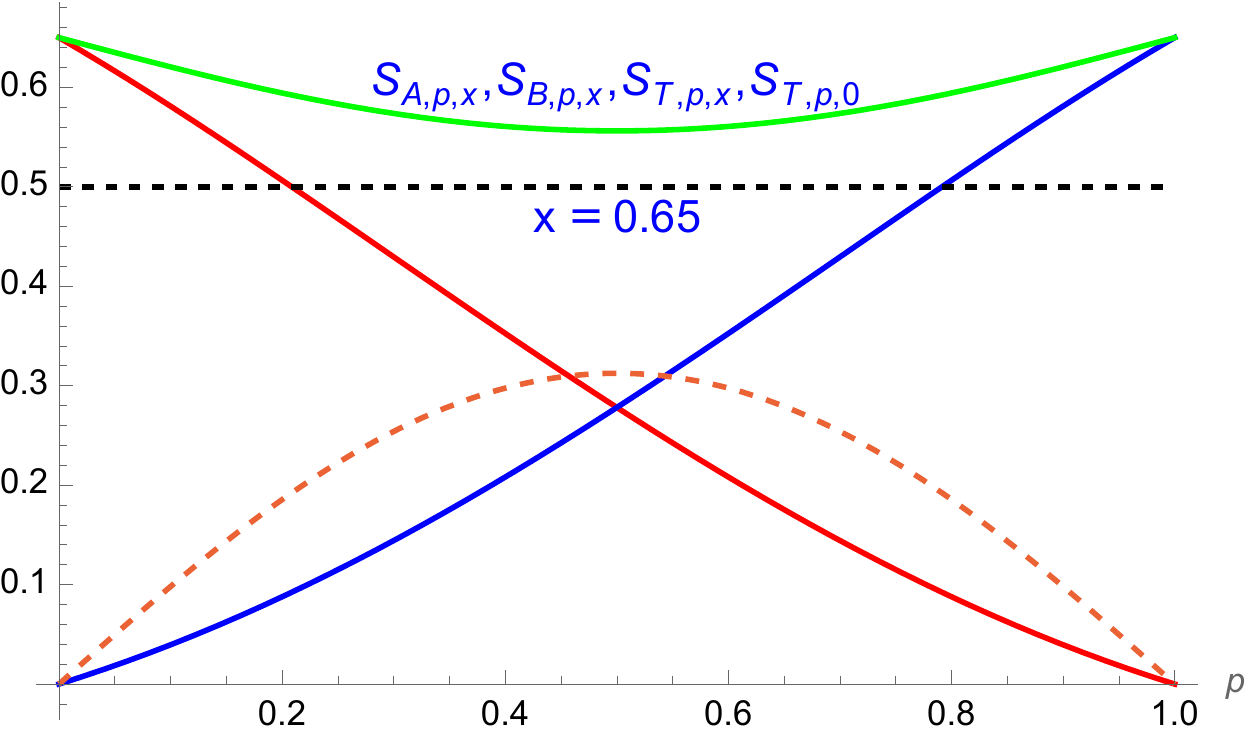}\\ \\ \\
\includegraphics[width=.50\textwidth]{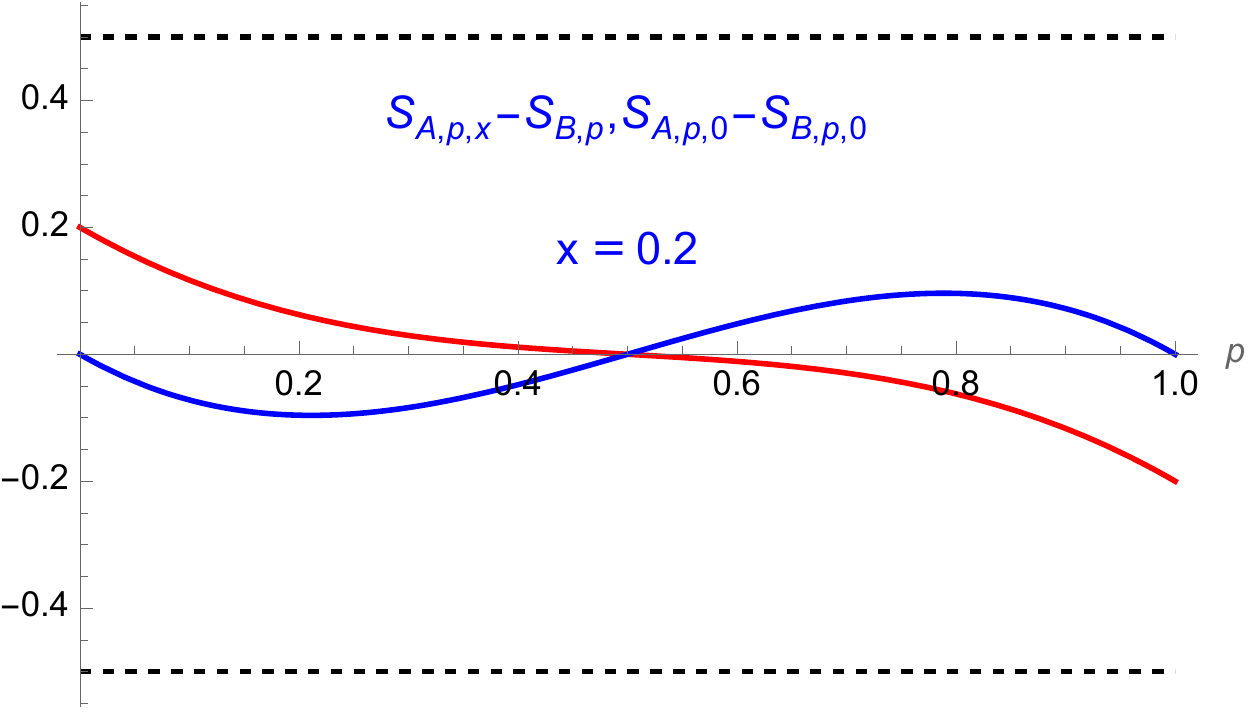}\quad
\includegraphics[width=.50\textwidth]{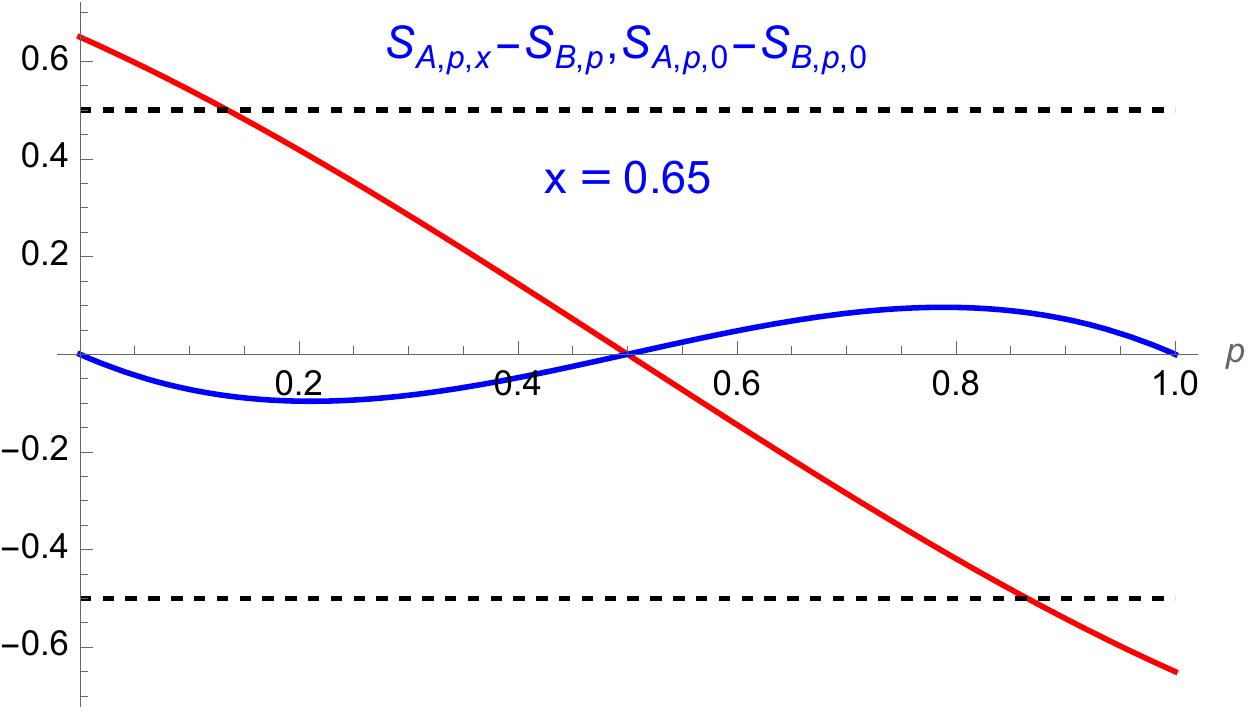}
\caption{Given $x=0.20$ (region 2, left part) and $x=0.65$ (region 3, right part), the Figure shows  the variations of $S_{A, p,x_-}, S_{A, p,x_+}, S_{A, p,x}, S_{A, p,0}$ (upper part), $S_{A, p,x}, S_{B, p,x}, S_{T, p,x}, S_{T, p,0}$ (middle part), $S_{A, p,x}- S_{B, p,x}, S_{A, p,0}-S_{B, p,0}$ (lower part), as a function of $p$.}
\label{sd}
\end{figure}

Fig.(\ref{sd}) shows the variations of theses quantities as a function of $p$ for $x=0.20$ and $x=0.65$, which are located respectively in Regions 2 and 3 where  $p_{c,x}=\frac{1}{2}$ is the attractor of the dynamics. 

\begin{itemize}
\item The top parts exhibits the magnitudes of the shifts with respect to opinion A where $S_{A, p,x_-}$ (in red) is the gain from local majority rule diminished by the contrarians. The gain from the loss of local majorities favorable to B is $S_{A, p,x_+}$ (in blue). The net gain for A is $S_{A, p,x}$ (in green). When only floaters are discussing, $S_{A, p,0}$ (in red dashed) is the gain for A.

\item The middle part exhibits the magnitudes of all shifts at the benefit of A ($S_{A, p,x}$ in red) and B ($S_{B, p,x}$ in blue) as well as the total shifts accounting for both A and B ($S_{T, p,x}=S_{A, p,x}+S_{A, p,x}$ in green). This total is also shown in the absence of contrarians with $S_{T, p,0}$ (in red dashed).

\item The bottom part exhibits the magnitude of the difference $S_{A, p,x}-S_{A, p,x}$ (in red)  in proportions of shifts at the benefit of respectively A and B. The equivalent $S_{A, p,0}-S_{A, p,0}$ (in blue) with only floaters is also shown.

\end{itemize}

To label the nature of the polarized state at the attractor, I evaluate the total proportion of shifts at $p=p_{c,x}=\frac{1}{2}$ as a function of $x$. From Eq.(\ref{sab4})  the associated proportions of individual shifts are $S_{A,\frac{1}{2},x}=S_{B,\frac{1}{2},x}=\frac{5+6 x }{32}$ and  $S_{T,\frac{1}{2},x}=\frac{5+6 x}{16}$. It yields 0.388 and 0.556 for respectively $x=0.20$ and $x=0.65$. The attractor is thus marked by a significant proportion of individual shifts between A and B. Despite the division of the community in two opposite halves, the high fluidity between the groups prevents the setting of hate between them. On this basis, I label that polarization as coexistence. The associated entropy is high.

It is worth to notice that having $S_{T,\frac{1}{2},0.65}=0.556$ points that above fifty percent of contrarians, the magnitude of individual shifts becomes smaller than the proportion of contrarians as expected with more contrarians than floaters and shown in the upper part of Fig. (\ref{sdd}).

\subsection{The magnitude of fluidity is a function of the discussing group size}

I showed in Section 3 that keeping invariant the tie configuration makes the update rule identical for both groups of four and three agents. However this equality does not hold with respect to the proportions of individual opinion shifts at the attractor $p_{c,x}=\frac{1}{2}$. Indeed, for groups of size three Eq.(\ref{sab4}) writes,

\begin{eqnarray}
S_{A, p, x_-}&=&(1 - x) p^2 (1 - p) , \nonumber\\
S_{A, p,x_+}&=&x  \left[(1 - p)^3 + 2 p (1 - p)^2\right] , \nonumber\\
S_{B, p,x_-}&=&(1 - x) p (1 - p)^2 , \nonumber\\
S_{B, p,x_+}&=&x  \left[p^3 + 2 p^2 (1 - p)\right] , 
\label{sab3}
\end{eqnarray}
yielding  $S_{T,\frac{1}{2},x}=\frac{1+2 x }{4}$ giving 0.350 and 0.575 for respectively $x=0.20$ and $x=0.65$.

\begin{figure} 
\centering
\includegraphics[width=0.50\textwidth]{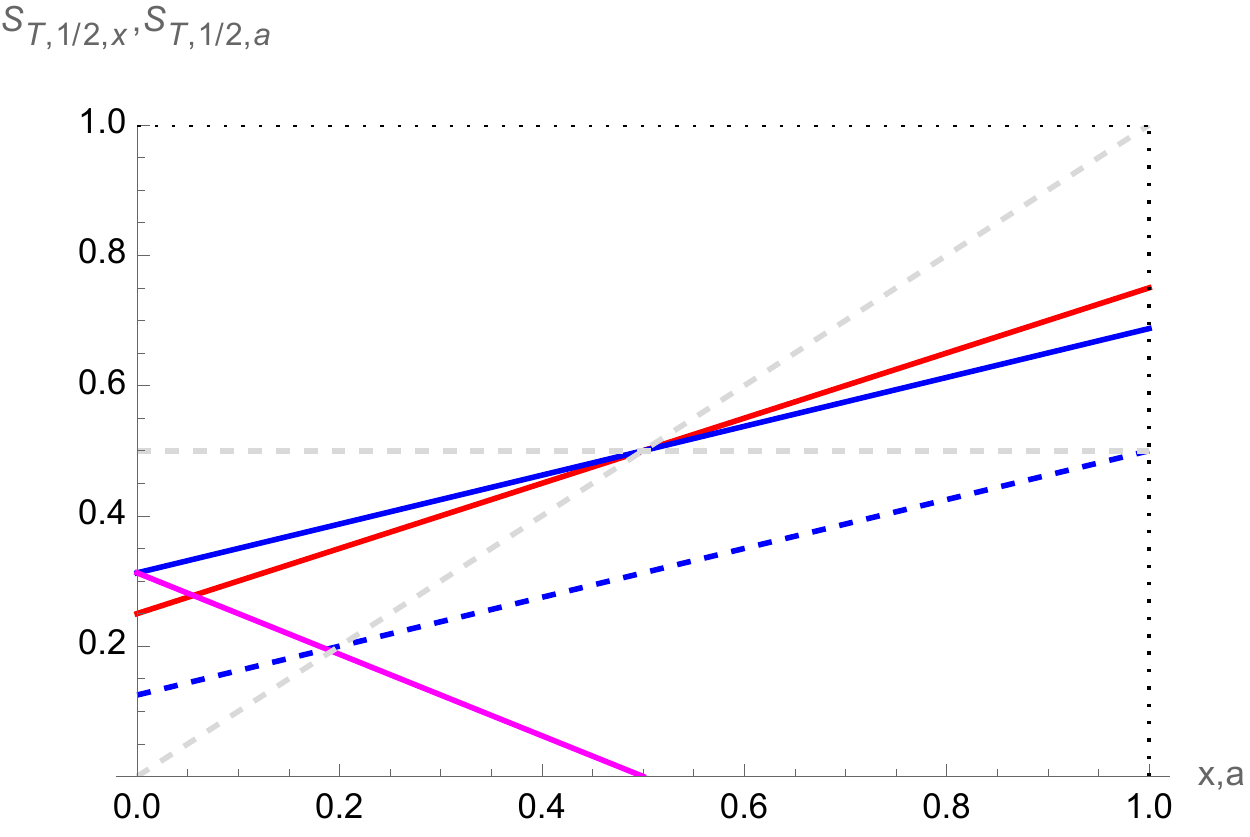} \\[1cm]
\includegraphics[width=.50\textwidth]{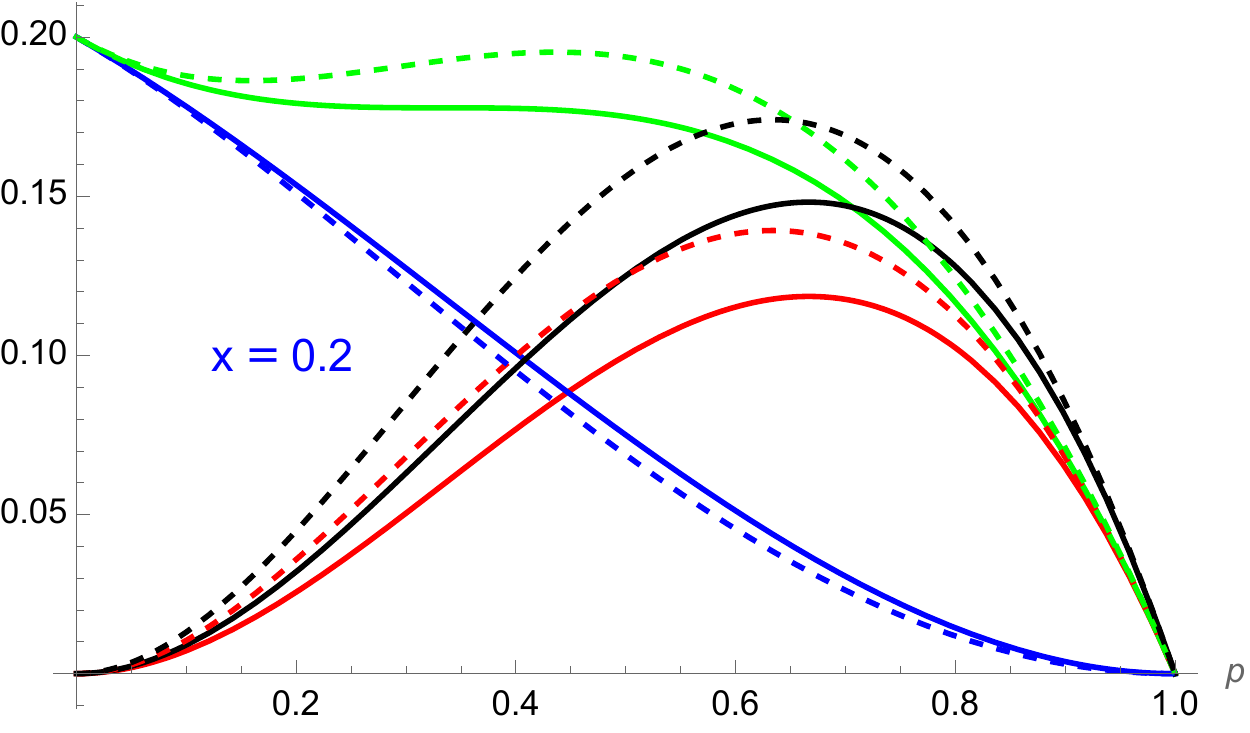}\hfill
\includegraphics[width=.50\textwidth]{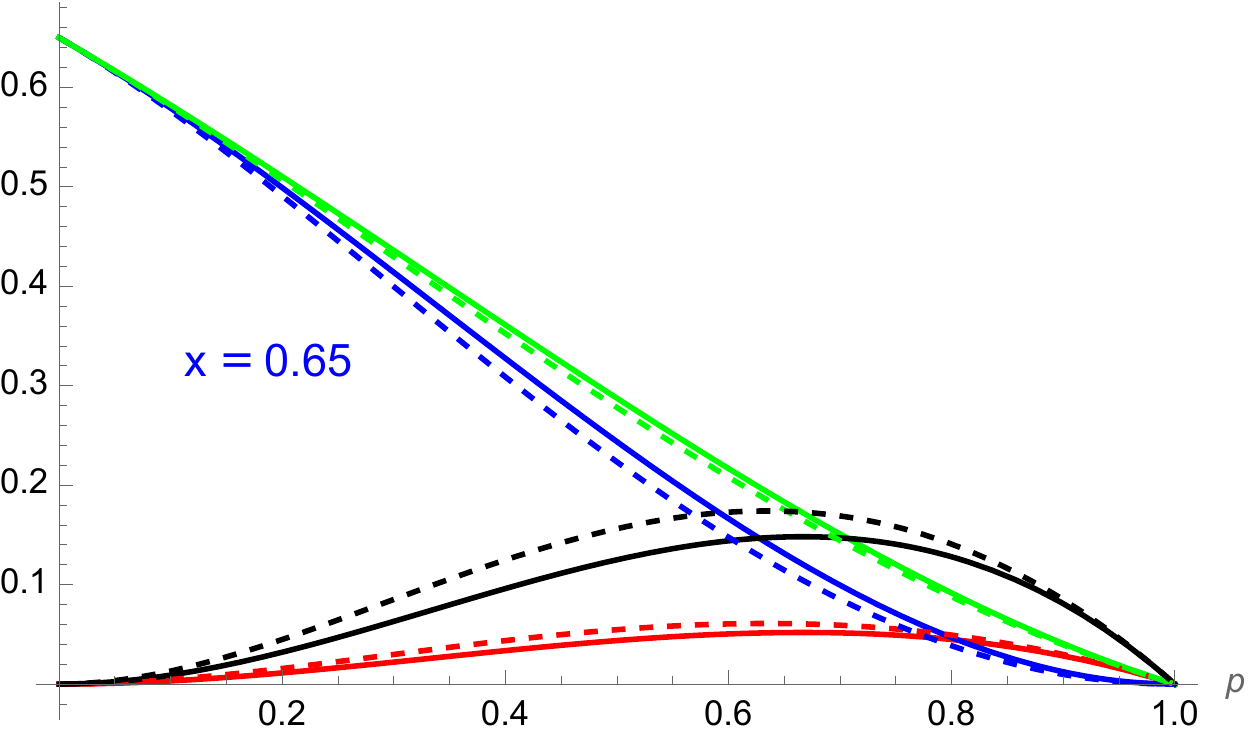} \\[1cm]
\includegraphics[width=.50\textwidth]{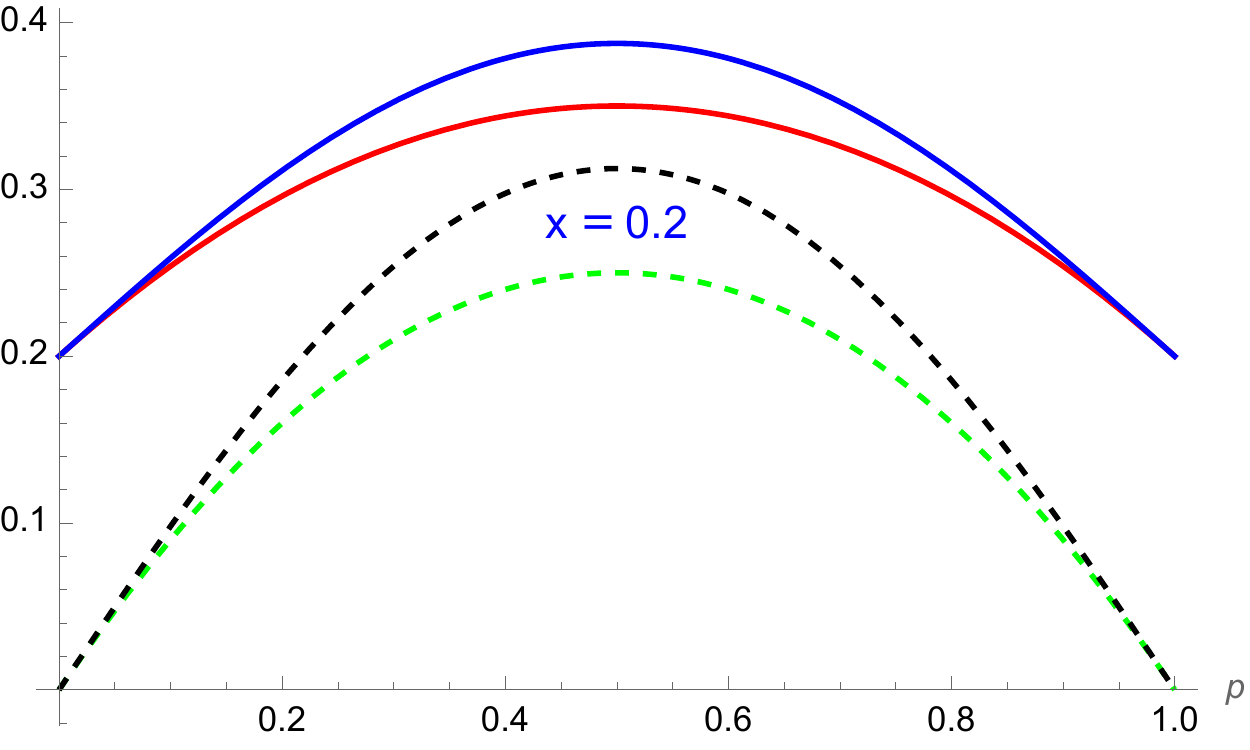}\hfill
\includegraphics[width=.50\textwidth]{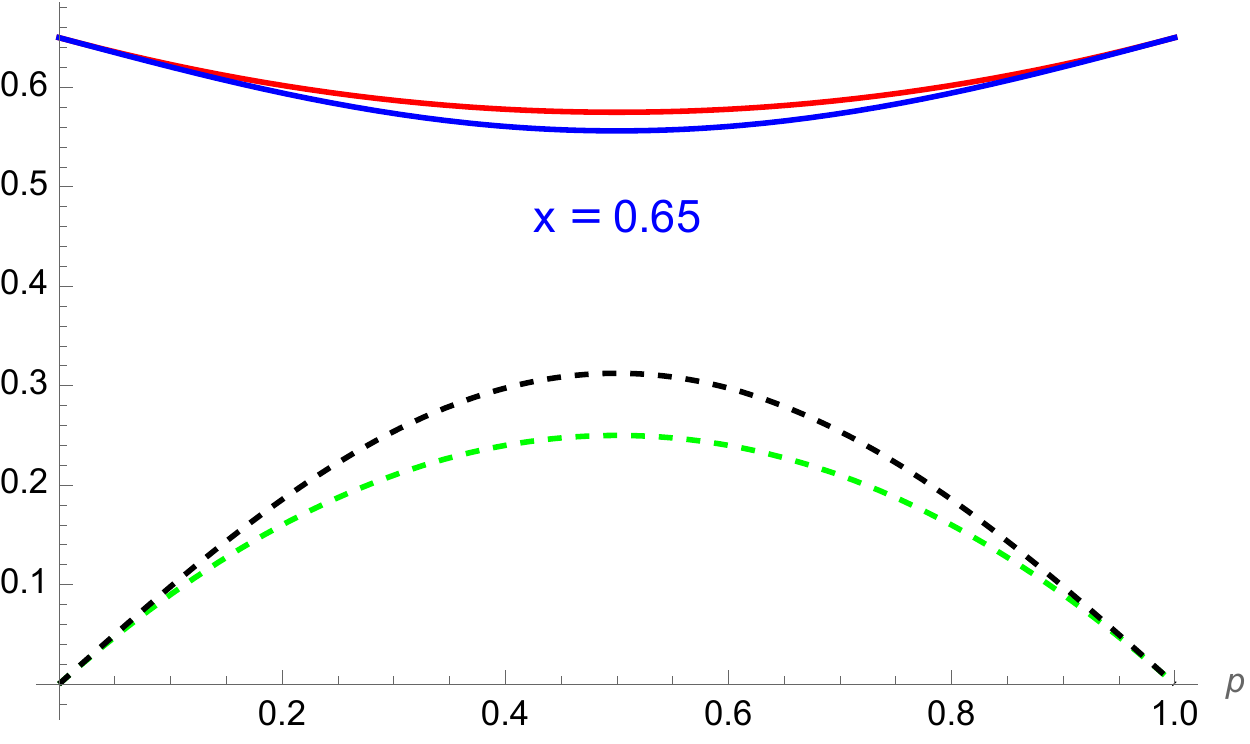}
\caption{The upper part shows $S_{T,\frac{1}{2},x}$ with $\frac{5+6 x}{16}$ for size 4 (in blue), $\frac{1+3x}{8}$ for size 4 with no change at ties (in dotted blue), $\frac{1+2 x}{4}$ for size 3 (in red) and $S_{T,\frac{1}{2},a}=\frac{5}{8} (\frac{1}{2}-x)$ for stubborns studied below (in magenta). The middle part shows $S_{A, p,x_-}, S_{A, p,x_+}, S_{A, p,x},$ for size 3 (in red, blue, green) and 4 (in dashed red, dashed blue, dashed green) at respectively  $x=0.20$ and $x=0.65$. The quantities at $x=0$ are added in both cases (in black for size 3 and dashed black for size 4).}
\label{sdd}
\end{figure}

Middle and lower parts of Fig. (\ref{sdd}) exhibit the proportions of individual shifts as a function of $p$ for groups of respectively size three and four with $x=0$, $x=0.20$ and $x=0.65$.

The results indicate that larger groups of discussion increase the magnitude of individual shifts and thus enlarge the coexistence part of the divided community.

However, it is worth to stress that in the case of size 4, keeping the tie 2A2B unchanged implies a weaker magnitude of individual shifts against turning the tie 2A2B into either 4A or 4B with equal probabilities $\frac{1}{2}$. In the first case, at the attractor, the magnitude of shifts is $\frac{1+3x}{8}$ versus $\frac{5+6x}{16}$ for the second case. For groups of size 3, the magnitude is  $\frac{1+2x}{4}$. Table (\ref{table}) shows the respective magnitudes as a function of $x$ and for $x=0.20$ and $x=0.65$. 

Thus, when nothing happens at a tie, groups of size 4 involve less shifts than groups of size three. Indeed, in the unchanged tie case, contrarians are neutralized at a tie by the absence of local majority. The proportion of groups at a tie being $6p^2(1-p)^2$, its value at the attractor $p=\frac{1}{2}$ is significant with $\frac{3}{8}=0.375$. 

Variations of the different magnitudes of individual shifts as a function of $x$ are shown in the upper part of Fig. (\ref{sdd}). All three cases exhibit a good deal of fluidity between the two opposite halves of the community. Contrarians add to the fluidity of the polarized state by their own shifts of opinion.
\renewcommand{\arraystretch}{2.5}
\begin{table}
\begin{center}
\begin{tabular}{|c|c|c|c|}
\hline
size & 3 & 4 & 4 unchanged \\
\hline 
$S_{T,\frac{1}{2},x}$  &$\frac{1+2x}{4}$ & $\frac{5+6x}{16}$ &  $\frac{1+3x}{8}$ \\ 
\hline 
$S_{T,\frac{1}{2},0.20}$  &  0.35 & 0.388 &  0.20 \\ 
\hline
$S_{T,\frac{1}{2},0.65}$  & 0.575 & 0.556 & 0.369 \\ 
\hline
\end{tabular}
\caption{Total proportion of individual shits $S_{T,\frac{1}{2},x}$ at the attractor $p_{c,x}=\frac{1}{2}$ for sizes 3, 4 with tie breaking and 4 with unchanged tie. The related values are also shown for $x=0.20$ and  $x =0.65$.}
\label{table}
\end{center}
\end{table}
\renewcommand{\arraystretch}{1}

\section{Stubbornness produces polarization}

Denoting $a$ and $b$ the respective proportions of stubborn agents along opinions A and B and keeping a balanced prejudice effect with $k=\frac{1}{2}$, update Eq. (\ref{p1}) writes \cite{inf},
\begin{equation}
\begin{split}
p_{1,a,b} =&p_0^4 + 4 p_0^3 (1 - p_0)+ 3 p_0^2 (1 - p_0)^2\\
+&\frac{a}{2}  (1 - p_0)^2 (2+p_0 ) -  \frac{b}{2} p_0^2([ 3 -p_0)  .
\end{split}
\label{p1s} 
\end{equation}

\begin{figure} [ht!]
\centering
\includegraphics[width=.50\textwidth]{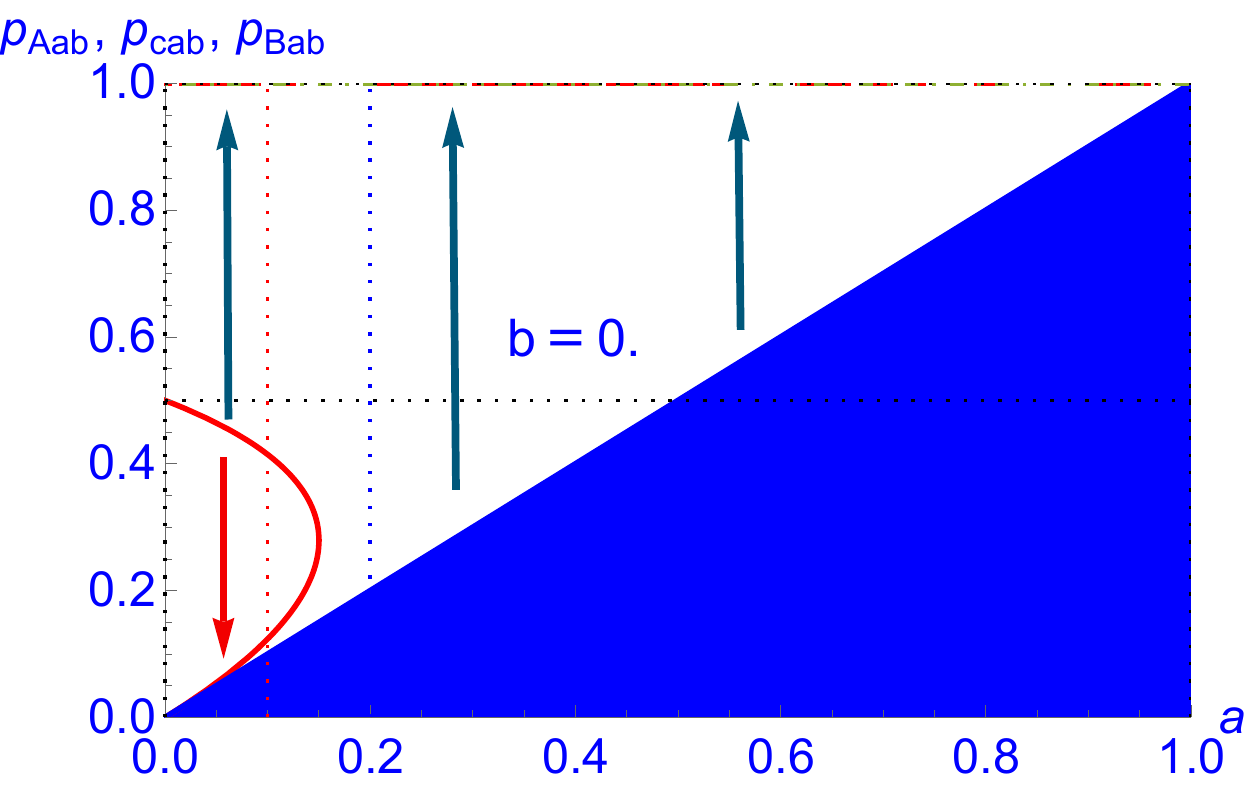}\hfill
\includegraphics[width=.50\textwidth]{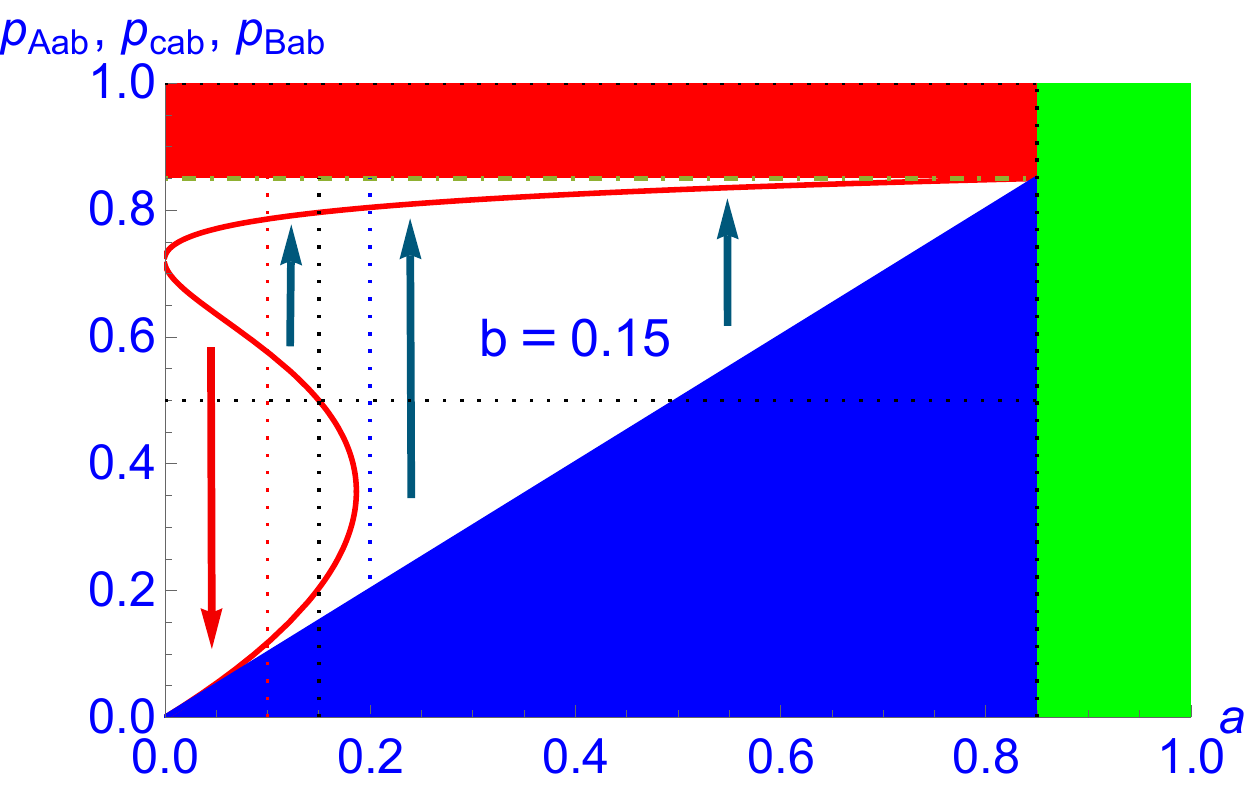} \\[1cm]
\includegraphics[width=.50\textwidth]{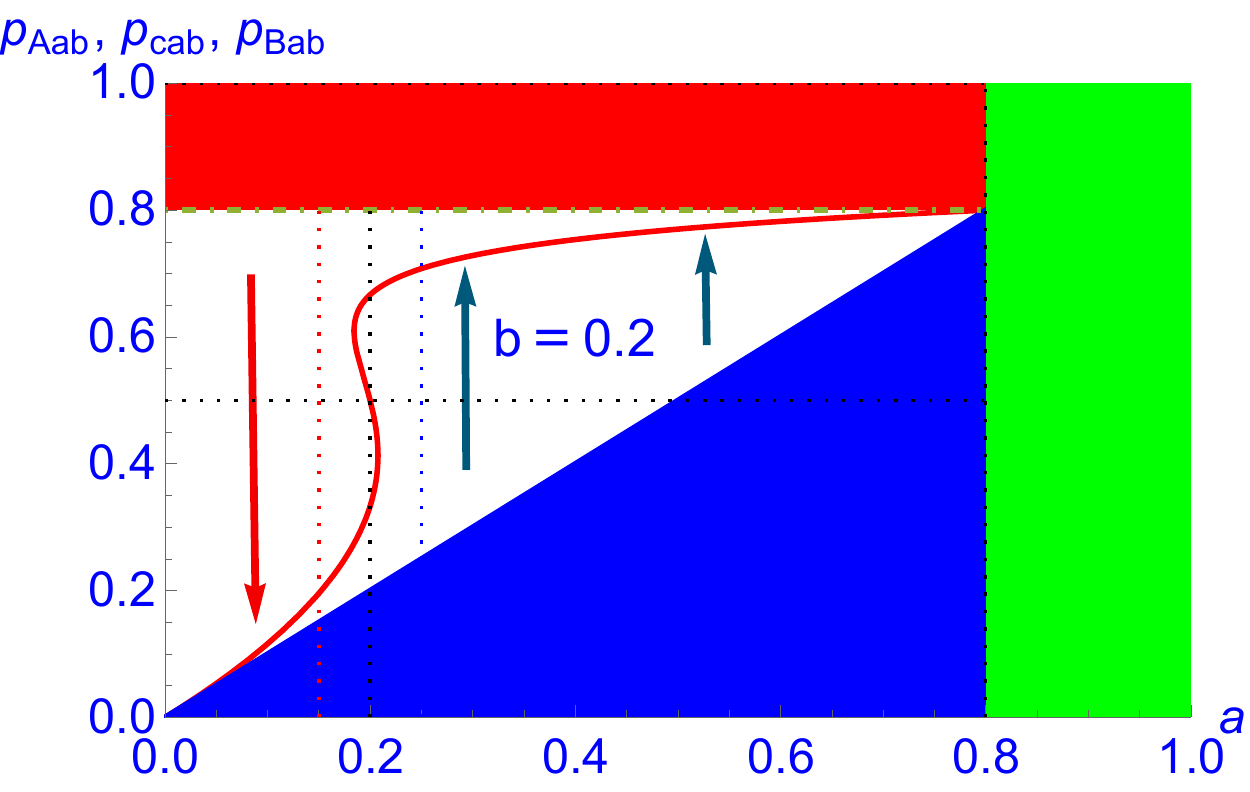}\hfill
\includegraphics[width=.50\textwidth]{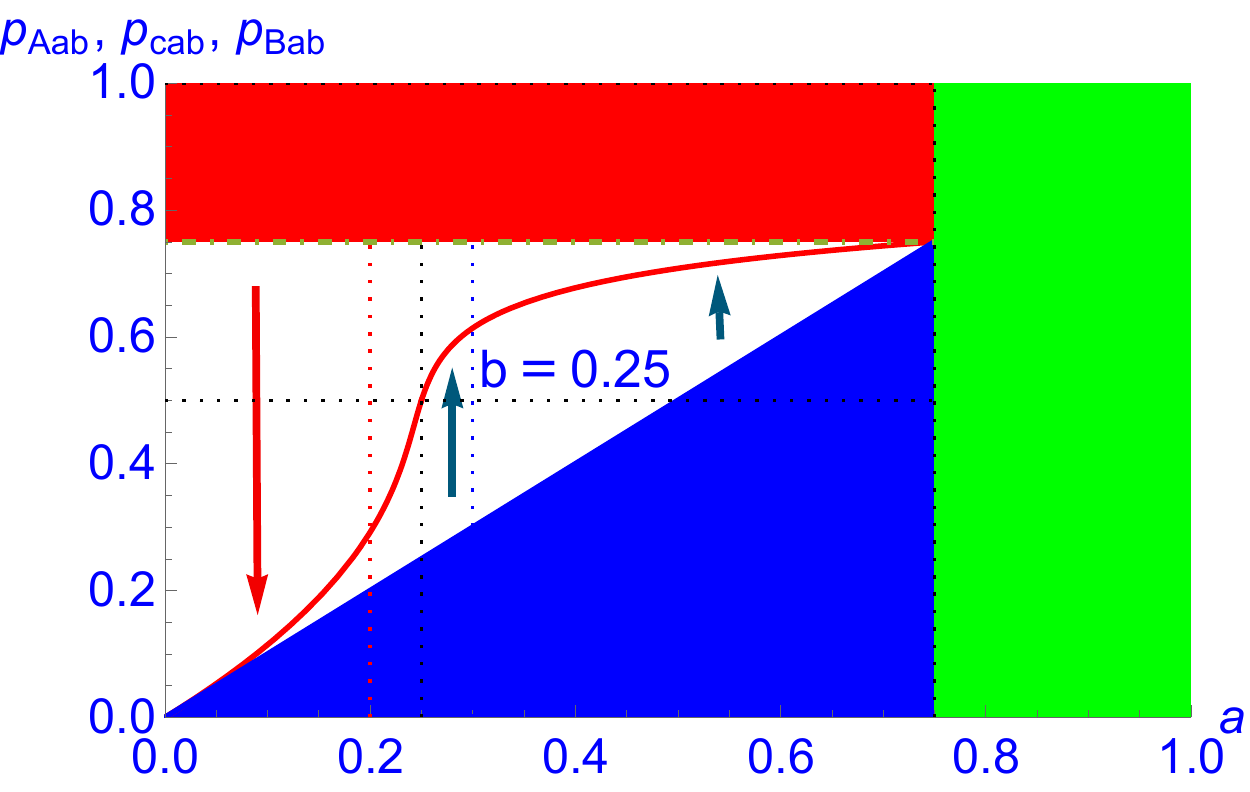}
\caption{Attractors and tipping points as a function of $a$ for a given $b$ from Eq. (\ref{p1s}). Four cases are exhibited with $b=0$, $b=0.15$, $b=0.20$ and $b=0.25$.}
\label{st}
\end{figure}

While Eqs. (\ref{p1},\ref{p1x}) have one parameter each, $k$ and $x$, Eq. (\ref{p1s}) has two parameters $a$ and $b$, which makes it more rich with respect to its dynamics. Fig. (\ref{st}) illustrates the various associated regimes exhibiting the interplay of attractors and tipping points as a function of $a$ for respectively $b=0, 0.15, 0.20, 0.25$. Two regimes are found.

\begin{description}
\item[Regime 1] The first regime is shown in Fig. (\ref{st}) for $b=0, 0.15, 0.20$ where two dynamics are taking place.  For small values of $a$ the dynamics is a tipping point dynamics but the associated region shrinks with increasing values of $b$. When $a$ get a bit large above about $0.20$ the dynamics becomes a single attractor dynamics with A always winning over B.

\item[Regime 2] The second regime shown in Fig. (\ref{st}) for $b=0.20$ has a unique type of dynamics with a single attractor dynamics. The opinion having more stubborns on its side eventually with over the other.
\end{description}

At this stage, for the sake of simplicity and without loss of generality I investigate further the case $a=b$, which allows an analytical solving. In this case, Eqs. (\ref{p1s}) becomes,
\begin{equation}
p_{1,a,b} =-2p_0^3+3p_0^2 + \frac{a}{2}  (2p_0^3-3p_0^2-3p+2) ,
\label{p1sa} 
\end{equation}
which yields three fixed points $p_{c,a}=\frac{1}{2}$ and,
\begin{equation}
p_{{A, a};{B,  b}}=\frac{-2 + a \pm \sqrt{4 -20 a+9 a^2}}{2 (-2 + a)} .
\label{pABx} 
\end{equation}
with last two being valid only in the range $0\leq a < a_c=\frac{2}{9} \approx 0.22$.

\begin{figure} [t]
\centering
\includegraphics[width=1\textwidth]{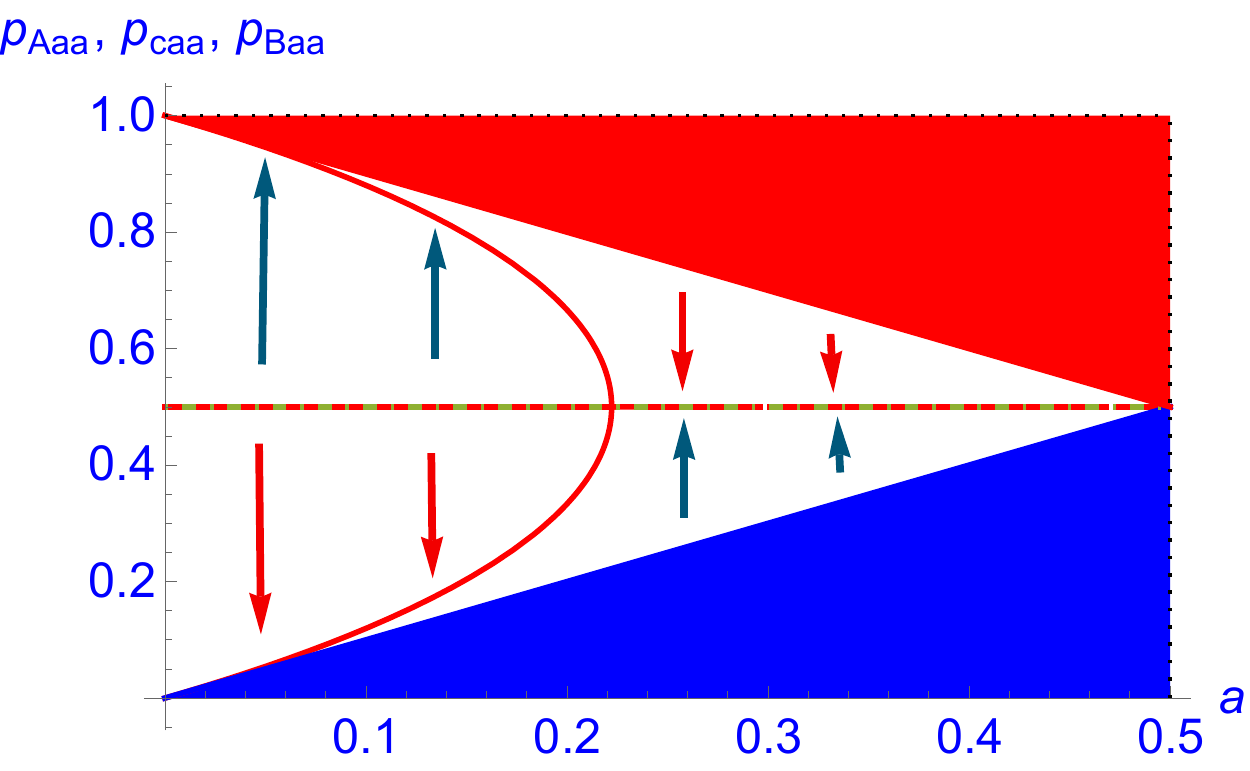}
\caption{The two distinct regimes produced by stubborn agents as a function of their proportion for $a=b$. A tipping point dynamics with $p_{c,x}=\frac{1}{2}$ prevails in the range $0\leq a < a_c$. The two associated attractors feature a stable coexistence of a majority and a minority. In the range $a_c \leq a \ \frac{1}{2}$ the dynamics is driven by one single attracto located at $\frac{1}{2}$. Any initial support $p_0<1-a$ moves monotonously towards $\frac{1}{2}$ with repeating local updates. There, both opinions coexist in a perfect balance.}
\label{sta}
\end{figure}

Therefore, regime 1 prevails in the range $0\leq a < a_c=\frac{2}{9}$ with $p_{c,a}=\frac{1}{2}$ acting as a tipping point between the two attractors $p_{A, a}$ and $p_{B, a}$. On the other hand, regime 2 prevails for $\frac{2}{9}\leq a \leq \frac{1}{2}$ with $p_{c,a}= \frac{1}{2}$ being the unique attractor as seen in Fig.(\ref{sta}).

The polarization arises in regime 2 monitored by the attractor $p_{c,a}= \frac{1}{2}$. It is worth to mention that in case $a\neq b$ with a small difference between $a$ and $b$ the polarization still occurs but then the two opposite parts are unequal. Either A ($a>b$) or B ($a<b$) has a numerical advantage at the attractor.

\subsection{Size three for the discussing group}
Earlier study of the case of size 3 with stubborn agents has the update Equation,
\begin{equation}
p_{1,a,b} =p_0^3 + 3 p_0^2(1 - p_0-\frac{b}{3})+  a (1 - p_0)^2
\label{pp1s} 
\end{equation}
instead of Eq. (\ref{p1s}). With $a=b$ the associated fixed points are $p_{c,a}=\frac{1}{2}$ (as for size 4) and,
\begin{equation}
p_{{A, a};{B,  b}}=\frac{1}{2} (1 \pm \sqrt{1-4a}) .
\label{pABx} 
\end{equation}
with last two being valid only in the range $0\leq a < a_c=\frac{1}{4} = 0.25$ instead of $a_c=\frac{2}{9} \approx 0.22$ for size 4.

It is worth to notice that while contrarians do not modify the value $x_c$ when moving from size 3 to 4, stubborns do modify $a_c$ from size 3 to 4.

\subsection{The rigidity of the stubborn made polarization}

Along the analysis of the fluidity of polarization produced by contrarians, I now evaluate the extend of freezing produced by stubborns within the two opposite halves of the population. Eqs.(\ref{sab4},\ref{sab3}) are deeply modified since stubbornness reduces only the shifting of agents from each opinion. The related shifts from B to A and from A to B write respectively,
\begin{eqnarray}
S_{A, p, a}&=&\frac{1}{2} p^2(1-p-b)(3-p) , \nonumber\\
S_{B, p, b}&= &\frac{1}{2} (1-p)^2(p-a)(2+p) ,
\label{stab4}
\end{eqnarray}
with by symmetry  $S_{A, p, a }=S_{B, 1-p, b}$.

Setting $a=b$, when $a>a_c=\frac{2}{9}$  the unique attractor of the dynamics is $p_{c,a}=\frac{1}{2}$ as seen in Fig. (\ref{sta}). The associated total proportion of shifts is $S_{T, p, a}=\frac{5}{8}(\frac{1}{2} - b)$ (Fig. (\ref{sdd})). It yields $0.188$ for $a=0.20$, which is quite low. Accordingly, at the attractor, the community is mostly frozen between two opposite halves. With $a=0.25$ the shifts amount to only $0.156$.

\section{Conclusion}

I have addressed the issue of polarization using the Galam model of opinion dynamics to find that within its frame, polarization arises spontaneously by the dynamics of opinions in the presence of either contrarians or stubborns. 

Moreover, by calculating the proportion of individual opinion shifts at the attractor $p_{c,a}=\frac{1}{2}$, I unveiled three types of polarization, which all arise from the same local majority update dynamics.

\begin{description}

\item[The fluid polarization] is produced by contrarian agents with a good deal of agents who keep shifting opinion between the two opposite parts of the community. This polarization do favor a coexistence between the group with a related high entropy.

\item[The frozen polarization] is produced by stubborn agents, which in turn provides a social and psychological basis for hate between the two split parts of a community with a kind of inside ignorance of the other side with  a low value entropy. The community is trapped into a rigid distribution of opinions.

\item[The segregated polarization] is produced by floaters In the absence of contrarians and stubborns. The dynamics leads toward unanimity within a connected social subgroup and to segregated polarization between adjacent not mixing sub-communities. Associated entropies are zero.

\end{description}

In a future work I intend to investigate the combined effect of mixing together the three kind of agents, the floaters, the contrarians and the stubborns.


\begin{thebibliography}{99.}


\bibitem {p1} L. G. Gajewski, J. Sienkiewicz and J. A. Holyst, Transitions between polarization and radicalization in a temporal bilayer echo-chamber model, Phys Rev E 105(2-1):024125 (2022)

\bibitem {p2} M. Kaufman, S. Kaufman and H. T. Diep, Statistical Mechanics of Political Polarization, Entropy 24,1262 (2022)

\bibitem {p3} P. T\"{o}rnberg, C. Andersson, K. Lindgren and S. Banisch, Modeling the emergence of affective polarization in the social media society, PLoS ONE 16(10): e0258259 (2021)

\bibitem {p4} A. Zafeiris, Opinion Polarization in Human Communities Can Emerge as a Natural Consequence of Beliefs Being Interrelated, Entropy 24, 1320 (2022)

\bibitem {p5} F. Baumann, P. Lorenz-Spreen, I. M. Sokolov, and M. Starnini, Emergence of polarized ideological opinions in multidimensional topic spaces, Physical Review X 11, 011012 (2021)

\bibitem {pp6} M. Doniec, A. Lipiecki, K. Sznajd-Weron, Consensus, Polarization and Hysteresis in the Three-State Noisy q-Voter Model with Bounded Confidence. Entropy 24,983 (2022)

\bibitem {pp7} P. Sobkowicz, Social Depolarization and Diversity of Opinions-Unified ABM Framework. Entropy 25, 568 (2023)


\bibitem {pa} https://en.wikipedia.org/wiki/January\_6\_United\_States\_Capitol\_attack

\bibitem {pb} https://en.wikipedia.org/wiki/2023\_Brazilian\_Congress\_attack

\bibitem {pc} https://en.wikipedia.org/wiki/2023\_Israeli\_judicial\_reform

\bibitem {paff2} S. Iyengar, Y. Lelkes, M. Levendusky, N. Malhotra and S. J. Westwood, The Origins and Consequences of Affective Polarization in the United States, Annu. Rev. Political Sci. 22, 129-46 (2019)


\bibitem {p6} S. Schweighofer, F. Schweitzer, and D. Garcia, A weighted balance model of opinion hyperpolarization, Journal of Artificial Societies and Social Simulation 23, 5 (2020)

\bibitem {p7}  N. Saintier, J. P. Pinasco and F. Vazquez, A model for the competition between political mono-polarization and bi-polarization, Chaos 30, 063146 (2020)

\bibitem {p8} D. Waldner and E. Lust, Unwelcome change: Coming to terms with democratic backsliding, Annual Review of Political Science 21, 93 (2018)



\bibitem{brazil}  R. Brazil, The physics of public opinion, Physics World,  January issue (2020)

\bibitem{frank} F. Schweitzer, Sociophysics, Physics Today 71, 40-47 (2018) 

\bibitem{book} S. Galam, Sociophysics: A Physicist's Modeling of Psycho-political Phenomena, 
Springer (2012)

\bibitem{bikas} B. K. Chakrabarti, A. Chakraborti and A. Chatterjee (Eds.),  Econophysics and Sociophysics: Trends and Perspectives, Wiley-VCH Verlag (2006)


\bibitem{nun1} M. G. E. da Luz, C. Anteneodo, N. Crokidakis and M. Perc, Sociophysics: Social collective behavior from the physics point of view, Chaos, Solitons and Fractals 170, 113379 (2023)




\bibitem{nun2}  N. Crokidakis and L. Sigaud, Role of inflexible minorities in the evolution of alcohol consumption, Journal of Statistical Mechanics: Theory and Experiment 9, 093403 (2022)

\bibitem{sen}  M. Tiwari, X. Yang and S. Sen, Modeling the nonlinear effects of opinion kinematics
in elections: A simple Ising model with random field based study, Physica A 582, 126287 (2021)

\bibitem{tot} G. T\`oth and S. Galam, Deviations from the Majority: A Local Flip Model, arXiv:2107.09344v1 (2021)

\bibitem{mala}  A. Kowalska-Stycze\' n and K. Malarz, Noise induced unanimity and disorder in
opinion formation, PLoS ONE 15(7): e0235313 (2020)

\bibitem{and}  M. V. Maciel, A. C. R. Martins, Ideologically motivated biases in a multiple issues opinion model,
Physica A 553 124293 (2020)

\bibitem{red} S. Redner, Reality-inspired voter models: A mini-review, Comptes Rendus Physique 20, 275-292 (2019) 

\bibitem {gm3} S. Galam and S. Moscovici, Towards a theory of collective phenomena. III: Conflicts and Forms of Power, European Journal of Social Psychology 25, 217-229 (1995)

\bibitem{kas1} A. Jedrzejewski, G. Marcjasz, P. R. Nail and K. Sznajd-Weron, Think then act or act then think?,
PLoS ONE 13(11):  e0206166 (2018)

\bibitem{bol} P. Singh, S. Sreenivasan, B. K. Szymanski, and G. Korniss, Competing effects of social balance and influence, Phys. Rev. E 93, 042306 (2016)

\bibitem{che} T. Cheon and J. Morimoto, Balancer effects in opinion dynamics, Phys. Lett. A 380, 429--434 (2016)

\bibitem{mau} S. Galam and A. Mauger, On reducing terrorism power: a hint from physics, Physica A: 323, 695-704 (2003)

\bibitem{nun3} N. Crokidakis, Radicalization phenomena: Phase transitions, extinction processes and control of violent activities, arXiv:2212.11361 (2022)

\bibitem{bag} F. Bagnoli and R. Rechtman, Bifurcations in models of a society of reasonable contrarians and conformists, Phys. Rev. E 92, 042913 (2015)

\bibitem{car} G. Carbone and I. Giannoccaro, Model of human collective decision-making in complex environments, Eur. Phys. J. B 88, 339 (2015)

\bibitem{kas2} K. Sznajd-Weron, J. Szwabi\'nski and R. Weron, Is the Person-Situation Debate Important for Agent-Based Modeling and Vice-Versa?, PLoS ONE  9: e112203 (2014)

\bibitem{zan} A. Chacoma and  D. H. Zanette, Critical phenomena in the spreading of opinion consensus and disagreement, Papers in Physics 6, 060003 (2014)

\bibitem{flo} R Florian and S Galam, Optimizing conflicts in the formation of strategic alliances, The European Physical Journal B-Condensed Matter and Complex Systems 16, 189-194 (2000)

\bibitem{mar} M. A. Javarone, Networks strategies in election campaigns, J. Stat. Mech. P08013 (2014)

\bibitem{igl} S. Goncalves, M. F.,Laguna and J. R. Iglesias, Why, when, and how fast innovations are adopted, Eur. Phys. J. B 85,
192 (2012) 

\bibitem{fas} A. Ellero, G. Fasano and A. Sorato, A modified Galam's model for word-of-mouth information exchange, Phys. A 388, 3901-3910 (2009) 


\bibitem{gim} M.C. Gimenez, L. Reinaudi and F. Vazquez, Contrarian Voter Model under the Influence of an Oscillating Propaganda: Consensus, Bimodal Behavior and Stochastic Resonance. Entropy 24,1140 (2022)  

\bibitem{iac} E. Iacominia and P. Vellucci, Contrarian effect in opinion forming: Insights from Greta Thunberg phenomenon, The Journal of Mathematical Sociology  47,123?169 (2023 )

\bibitem{bru} E. Brugnoli and M. Delmastro, Dynamics of (mis)information flow and engaging power of narratives, arXiv:2207.12264v2 (2022)



\bibitem{celia} A. M. Calv\~o, M. Ramos and C. Anteneodo, Role of the plurality rule in multiple choices, J. Stat. Mech. 023405 (2016)

\bibitem{mobilla} M. Mobilia, Fixation and polarization in a three-species opinion dynamics model, Eur. Phys. Lett.  95, 50002 (2011)



\bibitem{gtak} S. Galam and T. Cheon, Tipping points in opinion dynamics: a universal formula in five dimensions, Frontiers in physics, 8, 446 (2020)

\bibitem {uni2}  S. Galam, Opinion Dynamics and Unifying Principles: A Global Unifying Frame. Entropy 24, 1201 (2022)

\bibitem {vot} S. Galam,  Majority rule, hierarchical structures, and democratic totalitarianism: A statistical approach, J. of Math. Psychology 30, 426 (1986)

\bibitem {chop} S. Galam, B. Chopard, A. Masselot and M. Droz, Competing species dynamics: Qualitative advantage versus geography, Eur. Phys. J. B 4, 529-531 (1998) 

\bibitem{min} S. Galam,  Minority Opinion Spreading in Random Geometry,  Eur. Phys. J.  B 25 Rapid Note 403-406 (2002)

\bibitem{het} S. Galam, Heterogeneous beliefs, segregation, and extremism in the making of public opinions, Physical Review E 71 (4), 046123 (2005)

\bibitem{cont} S. Galam, Contrarian Deterministic Effects on Opinion Dynamics: The Hung Elections Scenario, Physica A 333, 453-460 (2004)

\bibitem {inf} S. Galam and F. Jacobs, The role of inflexible minorities in the breaking of democratic opinion dynamics, Physica A 381, 366-376 (2007)

\bibitem {pair} S. Galam, Collective beliefs versus individual inflexibility: The unavoidable biases of a public debate, Physica A: Statistical Mechanics and its Applications 390 (17), 3036-3054, (2011)

\bibitem {gmar} A. C. R. Martins and S. Galam, Building up of individual inflexibility in opinion dynamics, Physical Review E 87, 042807 (2013)


\end{thebibliography}
\end{document}